\documentclass[pra,groupedaddress,reprint,twocolumn,notitlepage]{revtex4}

\usepackage{amsfonts}
\usepackage{amsmath}
\usepackage{amssymb}
\usepackage{mathtools}
\usepackage{graphicx}
\usepackage{color}
\usepackage{mathrsfs}
\usepackage{bbm}
\usepackage[hidelinks]{hyperref}
\hypersetup{
     colorlinks   = true,
     citecolor    = blue,
     linkcolor    = blue
}
\usepackage{subfigure}
\usepackage{times,txfonts}
\usepackage{siunitx}

\makeatletter
\DeclareFontFamily{OMX}{MnSymbolE}{}
\DeclareSymbolFont{MnLargeSymbols}{OMX}{MnSymbolE}{m}{n}
\SetSymbolFont{MnLargeSymbols}{bold}{OMX}{MnSymbolE}{b}{n}
\DeclareFontShape{OMX}{MnSymbolE}{m}{n}{
    <-6>  MnSymbolE5
   <6-7>  MnSymbolE6
   <7-8>  MnSymbolE7
   <8-9>  MnSymbolE8
   <9-10> MnSymbolE9
  <10-12> MnSymbolE10
  <12->   MnSymbolE12
}{}
\DeclareFontShape{OMX}{MnSymbolE}{b}{n}{
    <-6>  MnSymbolE-Bold5
   <6-7>  MnSymbolE-Bold6
   <7-8>  MnSymbolE-Bold7
   <8-9>  MnSymbolE-Bold8
   <9-10> MnSymbolE-Bold9
  <10-12> MnSymbolE-Bold10
  <12->   MnSymbolE-Bold12
}{}

\let\llangle\@undefined
\let\rrangle\@undefined
\DeclareMathDelimiter{\llangle}{\mathopen}%
                     {MnLargeSymbols}{'164}{MnLargeSymbols}{'164}
\DeclareMathDelimiter{\rrangle}{\mathclose}%
                     {MnLargeSymbols}{'171}{MnLargeSymbols}{'171}
\makeatother

\DeclareMathOperator{\csch}{csch}

\newcommand{\ignore}[1]{}
\newcommand{\nobibentry}[1]{{\let\nocite\ignore\bibentry{#1}}}
\newcommand{\bibfnamefont}[1]{#1}
\newcommand{\bibnamefont}[1]{#1}

\newcommand{\ket}[1]{\left\vert#1\right\rangle}
\newcommand{\bra}[1]{\left\langle#1\right\vert}

\begin{document}

\title{Practical quantum metrology in noisy environments}

\author{Rosanna Nichols}
\affiliation{$\mbox{School of Mathematical Sciences, The University of Nottingham, University Park, Nottingham NG7 2RD, United Kingdom}$}
\author{Thomas R. Bromley}
\affiliation{$\mbox{School of Mathematical Sciences, The University of Nottingham, University Park, Nottingham NG7 2RD, United Kingdom}$}
\author{Luis A. Correa}
\affiliation{$\mbox{School of Mathematical Sciences, The University of Nottingham, University Park, Nottingham NG7 2RD, United Kingdom}$}
\author{Gerardo Adesso}
\affiliation{$\mbox{School of Mathematical Sciences, The University of Nottingham, University Park, Nottingham NG7 2RD, United Kingdom}$}
\email{gerardo.adesso@nottingham.ac.uk}

\begin{abstract}
The problem of estimating an unknown phase $ \varphi $ using two-level probes in the presence of unital phase-covariant noise and using finite resources is investigated. We introduce a simple model in which the phase-imprinting operation on the probes is realized by a unitary transformation with a randomly sampled generator. We determine the optimal phase sensitivity in a sequential estimation protocol, and derive a general (tight-fitting) lower bound. The sensitivity grows quadratically with the number of applications $ N $ of the phase-imprinting operation, then attains a maximum at some $ N_\text{opt} $, and eventually decays to zero. We provide an estimate of  $ N_\text{opt} $ in terms of accessible geometric properties of the noise and illustrate its usefulness as a guideline for optimizing the estimation protocol. The use of passive ancillas and of entangled probes in parallel to improve the phase sensitivity is also considered. We find that multi-probe entanglement may offer no practical advantage over single-probe coherence if the interrogation at the output is restricted to measuring local observables.
\end{abstract}

\date{\today}
\maketitle


\section{Introduction}

Advances in metrology are pivotal to improve measurement standards, to develop ultrasensitive technologies for defence and healthcare, and to push the boundaries of science, as demonstrated by the detection of gravitational waves \cite{LIGO}.
In a typical metrological setting, an unknown parameter $ \varphi $ is dynamically imprinted on a suitably prepared probe. We can think e.g.~of a two-level spin undergoing a unitary phase shift $ \hat{U} = \exp{\{-i\hat{H}\varphi\}} $. By subsequently interrogating the probe one builds an estimate $ \varphi_\text{est}$ for the parameter \cite{giovannetti2004quantum,PhysRevLett.96.010401,giovannetti2011advances}. The corresponding mean-square error $ \delta^2\varphi\equiv\langle(\varphi_\text{est}-\varphi)^2\rangle $ can be reduced, for instance, by using $ N $ uncorrelated identical probes. In that case, $ \delta^2\varphi $ scales asymptotically as $ 1/N $, which is referred to as the standard quantum limit  \cite{giovannetti2004quantum}. However, if those $ N $ probes were prepared in an entangled state, the resulting uncertainty could be  further reduced by an additional factor of $N$, leading to $ \delta^2\varphi \sim 1/N^2$. This ultimate quantum enhancement in resolution is termed Heisenberg limit and incarnates the {\it Holy Grail} of quantum metrology \cite{PhysRevLett.96.010401}.

In practice, the unitary dynamics of the probe will be distorted by noise, due to unavoidable interactions with its surroundings. Unfortunately, the metrological advantage of entangled probes over separable ones vanishes for most types of uncorrelated noise, such as spontaneous emission, depolarizing noise \cite{demkowicz2012elusive}, or phase damping \cite{PhysRevLett.79.3865,escher2011noisy}. Entanglement may remain advantageous though, provided one gains precise control over the noise strength, and only for limited cases such as time-inhomogeneous phase-covariant noise \cite{matsuzaki2011magnetic,PhysRevLett.109.233601,PhysRevA.92.010102,smirne2015ultimate,tanaka2015proposed}, transversal noise \cite{PhysRevLett.111.120401,PhysRevX.5.031010}, or when error-correction protocols may be used \cite{PhysRevLett.112.150802,PhysRevLett.112.080801}. Creating entangled states with a large number of particles is anyhow a costly process,  limited by technological constraints \cite{PhysRevLett.106.130506}. Furthermore, to fully harness the metrological power of entanglement in presence of noise, collective measurements on all $ N $ probes at the output would be generally required \cite{micadei}. This contrasts with the noiseless scenario, in which separable measurements (i.e., performed locally on each probe) suffice to attain the Heisenberg scaling \cite{PhysRevLett.96.010401}.

One can try to circumvent these problems by devising an alternative {\it sequential} or `multi-round' strategy, in which the parameter-imprinting unitary acts $ N $ consecutive times on a single probe before performing the final measurement. In absence of noise, this sequential setting is formally equivalent to the parallel one \cite{PhysRevA.88.042109}, the only difference being that quantum \textit{coherence} \cite{Baumgratz2014,Marvian2014,Marvian2016,Robus} takes over the instrumental role of entanglement. The sequential scheme seems more appealing from a practical viewpoint, as only a single probe needs to be addressed in both state preparation and final interrogation \cite{higgins2007entanglement}. However, the Heisenberg scaling of the precision cannot be maintained asymptotically in the sequential scenario either, once again due to the detrimental effects of noise.

Given the severe limitations that environmental disturbance places on quantum-enhanced metrology, for practical purposes it seems best to give up the prospect of super-classical {\it asymptotic} scaling of the resolution and to concentrate instead in using the {\it finite} resources available as efficiently as possible.

In this paper, we explore the optimization of phase estimation with a two-level probe, in the presence of \textit{unital phase-covariant} noise. To that end, in Sec.~\ref{sec:noise} we introduce a simple versatile model in which the noise is intrinsically accounted for: we take the generator $ \hat{H} $ of the phase shift to be partly unknown and sample instances of it from some probability distribution. The ensuing average mimics the environmental effects.
In Sec.~\ref{sec:sens} we calculate the {\it quantum Fisher information} (QFI) $F$ \cite{holevo2011probabilistic}, which can be meaningfully regarded as a quantitative benchmark for the optimal estimation sensitivity,  and derive a close-fitting lower bound $ f $ to it. Both quantities grow quadratically for small $ N $, reach a maximum at some $ N_\text{opt} $, and decay to zero as $ N $ increases further. In particular, we obtain $ N_\text{opt} $ from $f$ in terms of parameters directly accessible via process tomography, giving a useful prescription for practical phase estimation with a large guaranteed sensitivity. We do this for any unital phase-covariant qubit channel, hence delivering results widely applicable to a broad range of relevant physical processes, including those in which noise effects of the depolarizing type are dominant, such as spin-lattice relaxation at room temperature.

In Sec.~\ref{sec:ex} we then illustrate our results by choosing a specific distribution for the stochastic generator. We compare the QFI in the sequential setting (with and without passive correlated ancillas) with the actual phase sensitivity of given feasible measurements. For completeness, we also compute the QFI analytically in a parallel-entangled setting starting from an $ N $-qubit GHZ state. Although the QFI exhibits an asymptotic linear scaling in $N$ in such setting, we find that entangled probes may provide no practical advantage when their interrogation is restricted to measurements of local observables on each individual qubit. In fact, in such case the sensitivity for the parallel-entangled strategy reduces to that of the sequential one, where the `number of probes' comes to play the role of the `number of rounds'.
Our analysis, summarized in Sec.~\ref{sec:d}, reveals feasible solutions for quantum metrology based on the little-studied sequential paradigm (possibly supplemented by a passive ancilla), robust even under sizeable levels of noise.

\section{The noise model}\label{sec:noise}

\subsection{Uncertain phase generator}
Let us start by introducing our model for ease of illustration. In the sequential scenario, a two-level probe undergoes a sequence of $ N $ phase shifts $ \hat{U}_{\mathbf{n}} = \exp\{-i\varphi\hat{H}_{\mathbf{n}} \} $ before being interrogated. The generator can be written as $ \hat{H}_\mathbf{n} = \mathbf{n}\cdot\boldsymbol{\sigma}$, where the axis $ \mathbf{n}=( \sin{\theta}\cos{\phi},\,\sin{\theta}\sin{\phi},\,\cos{\theta} ) $ is sampled from some normalized probability distribution $ p(\theta,\phi) $
and  $ \boldsymbol{\sigma} $ is the vector of the three Pauli matrices. Thus the phase-imprinting operation $\Lambda_\varphi$, which transforms the probe state at each step, is
\begin{equation}
\hat{\varrho}_{N} = \Lambda_\varphi\,\hat{\varrho}_{N-1} \equiv \int_0^\pi\!\! d\theta\int_0^{2\pi}\!\! d\phi\, \hat{U}_\mathbf{n}\,\hat{\varrho}_{N-1}\,\hat{U}_{\mathbf{n}}^\dagger\, p(\theta,\phi)\,\sin{\theta}.
\label{eq:channel}
\end{equation}

The resulting qubit channel is completely positive, trace preserving,  {\it unital} (i.e.~$ \Lambda_\varphi\,\mathbbm{1} = \mathbbm{1} $), and contractive \cite{perez2006contractivity}, i.e.~any state will be asymptotically mapped to $ \hat{\varrho}_\infty\rightarrow\frac12\mathbbm{1} $; note also that Eq.~(\ref{eq:channel}) is akin to the classical simulation method of Ref.~\cite{demkowicz2012elusive}.
Without loss of generality we may take the average rotation axis $\langle\mathbf{n}\rangle = \int d\theta \int d\phi\,\mathbf{n}\,p(\theta,\phi) \sin \theta$ proportional to $(0,0,1)$, hence restricting to probability distributions $ p(\theta) $ with axial symmetry on the Bloch sphere, so that our qubit channel $\Lambda_{\varphi}$ is \emph{phase-covariant}, as it commutes with $\hat{H}_{\langle\mathbf{n}\rangle}$ \cite{smirne2015ultimate}.

Physically, we may think, for instance, of the free evolution of a nuclear spin with gyromagnetic ratio $ \gamma $ in an external magnetic field $ \mathbf{B} $ pointing along $z$. In that case $ \hat{H} = \frac{\gamma}{2} \hat\sigma_z$, so that $ \varphi = \frac{\gamma}{2} B\Delta t$, where $ \Delta t \gtrsim \frac{\hbar}{k_B T} $ is some coarse-grained time resolution, of the same order as the thermal fluctuations of the environment. The interactions with the surrounding nuclei at large temperatures result in random changes of the net direction of the magnetic field on our spin. This gives rise to a relaxation process towards the thermal equilibrium state $ \hat{\tau}_T\simeq\frac12\mathbbm{1} $. If the direction of $ \mathbf{B} $ were kept fixed and the environmental effects were accounted for by a fluctuating magnetic field intensity $ B $, the model would realize pure dephasing instead \cite{nielsen2000}, which is often dominant at short time scales.


\subsection{General unital phase-covariant qubit channels}

While the model in Eq.~(\ref{eq:channel}) can be conveniently adopted as a physical example to focus our analysis, the results derived in this paper will hold for a more general class of channels, namely all the unital phase-covariant qubit channels, whose description is recalled here.

Given a single-qubit state  $ \hat{\varrho} = \frac12\big(\mathbbm{1}+\mathbf{r}\cdot\boldsymbol{\sigma}\big) $, any single-qubit channel maps the Bloch vector $ \mathbf{r} $ into $ \mathbf{r}' = \mathbf{R}\,\mathbf{r} + \mathbf{t} $, where $ \mathbf{R} $ is a $ 3\times 3 $ real distortion matrix, and $ \mathbf{t} $ is a displacement vector.
For the most general unital phase-covariant channel \cite{bengtsson2006geometry}, $ \mathbf{t} = \mathbf{0}$  and
\begin{equation}
\mathbf{R}(\varphi)=\left(\begin{array}{ccc}
\lambda_{\perp}(\varphi)\cos{g(\varphi)} & - \lambda_\perp(\varphi)\sin{g(\varphi)} & 0 \\
\lambda_\perp(\varphi)\sin{g(\varphi)} & \lambda_\perp(\varphi)\cos{g(\varphi)} & 0 \\
0 & 0 & \lambda_\parallel(\varphi)
\end{array}\right).
\label{eq:phase_covariant}
\end{equation}
These channels encode information about $ \varphi $ not only in the rotation of the Bloch ball by a function $g(\varphi)$, but also in its deformation, through the  singular values of $\mathbf{R}(\varphi)$, namely $ \lambda_\parallel(\varphi) $ and the doubly-degenerate $ \lambda_\perp(\varphi) $, which must satisfy $ \lambda_\parallel \leq 1 $ and $ 2 \lambda_\perp \leq  1 + \lambda_\parallel $ (implying $ \lambda_\perp \leq 1 $) for the map to be completely positive \cite{braun2014universal}. Eq.~(\ref{eq:phase_covariant}) thus generalizes the canonical phase-covariant channels like phase damping, for which $g(\varphi)=\varphi$ \cite{smirne2015ultimate}.
In what follows we shall work with the $ 4 \times 4 $ Liouville representation $ \mathcal{K}(\varphi) $, which acts on vectorizations $ \left\vert\,\hat{\varrho}\right\rrangle \equiv \big( \bra{0}\hat\varrho\ket{0}, \bra{0}\hat\varrho\ket{1}, \bra{1}\hat\varrho\ket{0}, \bra{1}\hat\varrho\ket{1} \big)$ of any density matrix  $ \hat\varrho $ in the computational basis as  $\left\vert\,\hat{\varrho}\right\rrangle \mapsto \mathcal{K}(\varphi) \left\vert\,\hat{\varrho}\right\rrangle $, where
$ \mathcal{K}(\varphi) $ writes as
\begin{equation}
\mathcal{K}(\varphi) = \frac12\left(\begin{matrix}
1+\lambda_\parallel(\varphi) & 0 & 0 & 1-\lambda_\parallel(\varphi) \\
0 & 2\lambda_\perp(\varphi) e^{-i g(\varphi)} & 0 & 0 \\
0 & 0 & 2\lambda_\perp(\varphi) e^{i g(\varphi)} & 0 \\
1-\lambda_\parallel(\varphi) & 0 & 0 & 1+\lambda_\parallel(\varphi)
\end{matrix}\right).
\label{eq:liouville_K}
\end{equation}

\section{Phase sensitivity}\label{sec:sens}

\subsection{Quantum Fisher information}

 We will now analyze the sensitivity for estimating a phase imprinted by any unital phase-covariant qubit channel, starting with the sequential estimation setting. Recall that the estimate $ \varphi_\text{est} $ is built by measuring some observable $ \hat{O} $ on the final state of the probe $\hat{\varrho}_{N}$ after $N$ successive applications of the channel. We can then gauge the phase sensitivity of $ \hat{O} $ as
\begin{equation}\label{eq:cfi_general}
  F^{\hat{O}} \equiv \frac{\partial_\varphi\langle\hat{O}\rangle^2}{\Delta^2 \hat{O}},
\end{equation}
  where $ \langle\hat{O}\rangle \equiv \text{tr}\{ \hat{\varrho}_N\hat{O} \} $ and $ \Delta^2\hat{O} \equiv \langle\hat{O}^2\rangle - \langle\hat{O}\rangle^2 $ \cite{PhysRevLett.72.3439}. In general, $F^{\hat{O}} \leq I^{\hat{O}}$, where $I^{\hat{O}}$ is the   {\it classical} Fisher information of a projective measurement  onto the eigenstates $\{ \ket{o_i}\} $ of the observable $\hat{O}$ (assumed non-degenerate) \cite{PhysRevLett.72.3439}. In particular, one can verify that equality holds for all the explicit examples presented later in the paper, henceforth we shall simply refer to $F^{\hat{O}}$ as the phase sensitivity associated to the (generally suboptimal) observable ${\hat{O}}$.

  The QFI $ F $, which captures geometrically the rate of change of the evolved probe state $\hat{\varrho}_N$ under an infinitesimal variation of the parameter $\varphi$,  corresponds to the classical Fisher information of an \textit{optimal} observable (i.e. $F  = \sup_{\hat{O}} I^{\hat{O}} $) and is thus a meaningful benchmark for the best estimation protocol. Such an optimal observable is diagonal in the eigenbasis of the symmetric logarithmic derivative (SLD) $\hat{L}$, defined implicitly by \cite{barndorff2000fisher}
 \begin{equation}
 \hat{L}\,\hat{\varrho}_N + \hat{\varrho}_N\hat{L} \equiv 2 \partial_\varphi\,\hat{\varrho}_N .
 \end{equation}

Note that the prominent role of the QFI $F$ in quantum estimation theory is well established in an asymptotic setting, by virtue of its appearance in the quantum Cram\'er-Rao bound, \cite{holevo2011probabilistic}, $\delta^2\varphi \geq 1/(M F)$, which becomes tight in the limit of a large number $M \gg 1$ of independent repetitions. However, the QFI also enters in both the van Trees inequality \cite{gilltrees} and the Ziv-Zakai bound \cite{mankey}, which can provide tighter and more versatile  bounds on the mean-square error $\delta^2\varphi$ in the relevant case of finite $M$ (including Bayesian settings). Therefore, we shall adopt the QFI as our main figure of merit, in keeping with the quantum metrology bulk literature  (see e.g.~\cite{kolo2016} for a recent discussion) and in compliance with the spirit of this paper, which focuses on the use of finite resources to retain quantum enhancements in the estimation sensitivity.
 We will also assume maximization of $ F $ over the initial state of the probe unless stated otherwise.

To calculate the QFI $F_N(\varphi)$ we make use of the formula
\begin{equation}\label{eq:qfi_formula_F_N}
F_N(\varphi) = 4 \sum_{i,j} \frac{q_i}{(q_i+q_j)^2} \vert\langle\psi_i\vert\partial_\varphi\,\hat{\varrho}_N\vert\psi_j\rangle \vert^2,
 \end{equation} where $ \ket{\psi_i} $ and $ q_i $ are, respectively, the eigenvectors and eigenvalues of $ \hat{\varrho}_N $ \cite{paris2009quantum,0253-6102-61-1-08} (excluding terms with $ q_i + q_j = 0 $ from the sum), and we have made the dependence of the QFI on $\varphi$ and $N$ explicit. By preparing the probe in the optimal, maximally coherent state $ \ket{+} = \left(\ket{0}+\ket{1}\right)/\sqrt2 $ \cite{PhysRevA.88.042109}, we obtain exactly
\begin{equation}
F_N(\varphi) = N^2\left[\lambda_\perp^{2N}(\partial_\varphi g)^2+\frac{(\partial_\varphi\lambda_\perp)^2\lambda_\perp^{2N-2}}{1-\lambda_\perp^{2N}}\right] . 
\label{eq:qfi_formula_general}
\end{equation}

The QFI thus grows as $ \sim N^2 $ for small $N$, reaches a peak at some $ N_\text{opt} $, and then decays asymptotically to zero. A similar qualitative behaviour has been recently reported under other types of (non-unital) noise, such as erasure, spontaneous emission, and phase damping \cite{demkowicz2014using,yousefjani2016framework}.

\subsection{Optimal `sampling time'}

\begin{figure}
\includegraphics[width=\columnwidth]{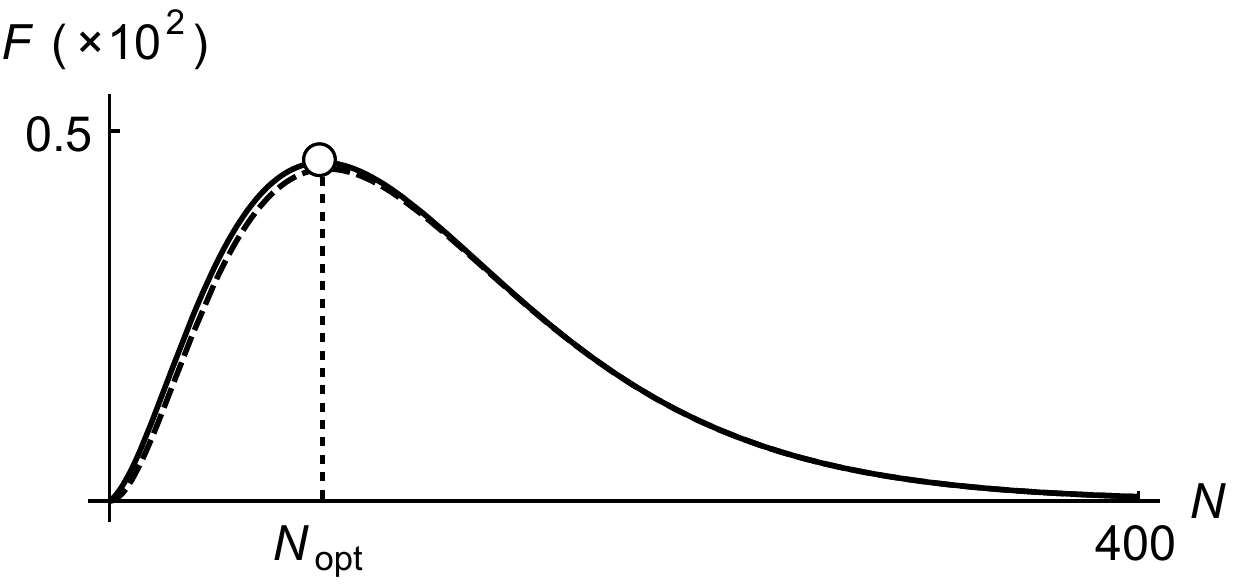}
	\caption{Quantum Fisher information $ F_N(\varphi) $ (solid) and lower bound $ f_N(\varphi) $ (dashed) versus number of rounds $ N $ under the channel of Eq.~\eqref{eq:channel} for the von Mises-Fisher distribution $ p_\kappa(\theta) $, with $\varphi=0.1$ and $\kappa=1$, and probe initialized in the $\ket{+}$ state.
The maximum of $ f_N(\varphi) $ is at $ N'_\text{opt} = 85 $, only one round further  the actual maximum of  $ F_N(\varphi) $. $ N $ is discrete and the continuous lines are a guide to the eye. All the quantities plotted are dimensionless.}
\label{fig1}
\end{figure}

In order to optimize our sequential protocol, we need a practical way to determine $ N_\text{opt} $. To that end, we aim to bound the QFI from below. In general, one may use $\max_{\hat{\varrho}}\llangle\hat{\varrho}\vert \big(\partial_\varphi\mathcal{K}^N\big)^\dagger \partial_\varphi\mathcal{K}^N \vert\hat{\varrho}\rrangle \leq F_N(\varphi)$ \cite{alipour2014quantum,yousefjani2016framework}, which is not necessarily tight. Luckily, for the family of channels encompassed by our $ \Lambda_\varphi $, a tighter bound $f_N(\varphi)$ stemming from $ \vert\vert \big(\partial_\varphi\mathcal{K}^N\big)^\dagger \partial_\varphi\mathcal{K}^N \vert\vert $, where $ \vert\vert\cdots\vert\vert $ stands for the operator norm, can be established  (see Appendix~\ref{app:bound}). Concretely, we will thus define
\begin{equation}
f_N(\varphi) \equiv N^2\, \lambda_\perp^{2N-2} [\lambda_\perp^2 (\partial_\varphi g)^2+(\partial_\varphi\lambda_\perp)^2] \leq F_N(\varphi),
\label{eq:bound_result}
\end{equation}
which closely follows the behaviour of the QFI, as confirmed by our numerical analysis (cf.~Fig.~\ref{fig1}). Maximizing $ f_N(\varphi) $ yields $ N'_\text{opt} = -1/ \log{\lambda_\perp} \simeq N_\text{opt}$ (rounded to the nearest integer and using natural logarithm). This approximation to the optimal `sampling time' $ N_\text{opt} $ only depends on $ \lambda_\perp $, which is experimentally accessible through process tomography \cite{karpinski2008fiber}.

It follows that there exist observables with a phase sensitivity of \textit{at least} $ f_N(\varphi) $, which could be made grow above $ f_{N'_\text{opt}} = (\partial_\varphi\lambda_\perp/e\lambda_\perp\log{\lambda_\perp})^2 $ by just running the sequential protocol for $ N'_\text{opt} $ iterations. Once the precision has been optimized at the single-probe level, it can be scaled-up ``classically'' by increasing the number of independent probes. We emphasize that Eqs.~\eqref{eq:qfi_formula_general} and \eqref{eq:bound_result} apply \textit{generally} to any phase-covariant channel preserving the identity. Therefore, we have provided useful guidelines for precise phase estimation under a wide class of physically motivated noise models.

\section{Example: ``Gaussian'' depolarizing noise}\label{sec:ex}

\subsection{Sequential setting}

To illustrate our results, we shall pick the von Mises-Fisher distribution \cite{fisher1953dispersion} $ p_\kappa(\theta)=\kappa e^{\kappa\cos{\theta}}/(4\pi\sinh{\kappa}) $ for our random generator $ \hat{U}_\mathbf{n} $ in Eq.~\eqref{eq:channel}. This distribution can be seen as the counterpart of a Gaussian over the Bloch sphere: it becomes uniform for $ \kappa\rightarrow 0 $ (i.e. $ \mathbf{n} $ is equally likely to point in any direction, as in a black-box scenario \cite{interpower,avsk}), whereas it localizes sharply around the $ z $--axis for $ \kappa \rightarrow \infty $. In Appendices \ref{app:kraus} and \ref{app:sequential}, we provide an explicit operator-sum representation for this choice of $ \Lambda_\varphi $, as well as expressions for the resultant $ \mathcal{K}(\varphi) $, $ F_N(\varphi) $, and $ f_N(\varphi) $.

Recall that saturating the precision bound set by the QFI requires interrogating the probe in the eigenbasis of the SLD. However, the SLD basis usually depends explicitly on the unknown parameter $ \varphi $ (and, in this case, on the noise parameter $\kappa$) so that, in practice, one would need to implement adaptive feedforward estimation procedures \cite{paris2009quantum}, or could have to resort to sub-optimal phase-independent estimators $ \hat{O} $. It is therefore particularly interesting to see how the sensitivity of {\it accessible} observables compares with the QFI. We can study, for instance, the phase sensitivity of $ \left\langle\hat{\sigma}_x\right\rangle = \text{tr}\big\{ \hat{\sigma}_x\,\hat{\varrho}_N \big\} $ or, equivalently, that of $ \mathbf{m}\cdot\boldsymbol{\sigma} $ for any $ \mathbf{m} $ in the equatorial plane. Explicit formulas are provided in Appendix~\ref{app:sequential}, and the resulting $ F^{\hat{\sigma}_x} $ is plotted alongside $ F_N(\varphi) $ in Fig.~\ref{fig2} (dashed and solid curves, respectively).

Interestingly, as $ N $ increases, the sensitivity of $ \left\langle\hat{\sigma}_x\right\rangle $ is seen to {\it oscillate} regularly between zero and the QFI. This can be intuitively understood if one thinks of the `dynamics' of $ \hat{\varrho}_N $ in the Bloch sphere: The maximally coherent preparation $ \hat{\varrho}_0 = \ket{+}\bra{+} $ lies along the equator, on the Bloch sphere surface. Then, the iterative application of  $ \Lambda_\varphi $ gives rise to a trajectory which inspirals on the equatorial plane as $ \hat{\varrho}_N $ approaches its fixed point $ \hat{\varrho}_\infty = \frac12\mathbbm{1} $ at the center of the sphere. This is just a combination of the unitary rotation around the $ z $--axis and the loss of purity that results from the average in Eq.~\eqref{eq:channel}. The eigenstates of the SLD follow the rotation of $ \hat{\varrho}_N $, whereas our actual measurement basis remains fixed. As a result, the sensitivity of $ \hat{\sigma}_x $ oscillates between $ 0 $ and $ F_N(\varphi) $, with the latter saturated when the two bases coincide  (see Appendix~\ref{app:sequential} for a visual description).

\subsection{Parallel-entangled setting}

As a comparison, let us now consider the parallel-entangled strategy, i.e., a single-round protocol starting from an entangled state of the $ N $ qubits. We will choose the maximally-entangled GHZ state $ \frac{1}{\sqrt{2}}(\ket{0}^{\otimes N} + \ket{1}^{\otimes N}) $ \cite{greenberger1989going} as initial preparation. Although this may not be optimal for noisy parameter estimation \cite{PhysRevLett.79.3865,escher2011noisy,metromps}, it comes with the advantage that it keeps a simple form under the local application of our channel on all $ N $ probes. It also has the same degree of quantum coherence as the single-qubit $ \ket{+} $ state, as measured by the $ \ell_1 $--norm \cite{Baumgratz2014}. This choice allows us to obtain an analytic expression for the QFI (see Appendix~\ref{app:parallel}), which is plotted in Fig.~\ref{fig2} (dotted curve) along with the QFI of the sequential case. Note that, while we use the same notation `$ N $' for the number of rounds in the sequential case and for the number of probes in the parallel one, these are essentially different resources. One can nonetheless make sense of the comparison between the two metrological settings by invoking their formal equivalence in absence of noise \cite{PhysRevA.88.042109}, and by recalling that $N$ equals the overall number of interactions with the phase-imprinting channel in both cases.
\begin{figure}
\includegraphics[width=\columnwidth]{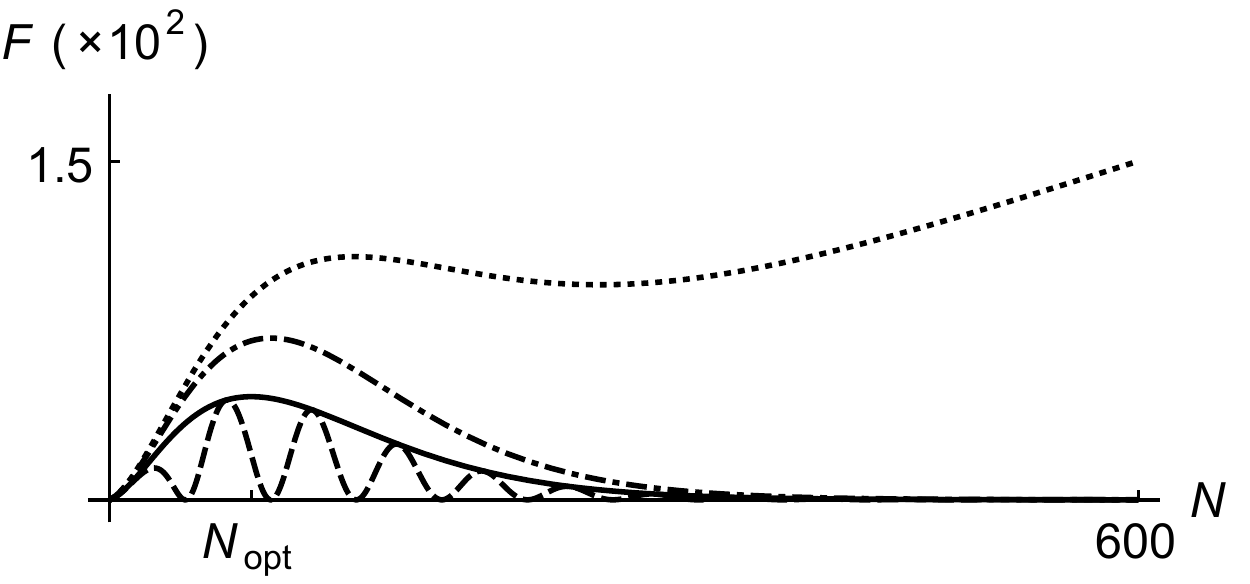}
\caption{Quantum Fisher information versus $N$ for: the $N$-round sequential setting with a single-qubit initial state $\ket{+}$ (solid), the $N$-round sequential setting with a passive ancilla and two-qubit initial Bell state (dot-dashed), and the parallel-entangled setting with $N$-qubit initial GHZ state (dotted). The dashed line amounts to the phase sensitivity of $\hat{\sigma}_x$, $\hat{\sigma}_x^{\otimes 2}$, and $\hat{\sigma}_x^{\otimes N}$ in each of the three settings, respectively. The model parameters are the same as in Fig.~\ref{fig1}. All the quantities plotted are dimensionless.}
\label{fig2}
\end{figure}

The resulting parallel QFI exhibits a linear asymptotic scaling with $N \gg 1$, unlike the sequential setting. However, even if such a large $N$-qubit entangled probe could be prepared, its maximum sensitivity would only be saturated by some phase-dependent \textit{collective} measurement on all $ N $ probes \cite{micadei}.
Indeed, in Appendix~\ref{app:parallel} we show that this parallel QFI  may be split into two contributions: one, with a profile similar to that of the sequential $ F_N(\varphi) $, stemming from the matrix elements of the output state in the subspace of total angular momentum $ J^2 = N/2 $, and another one, related to the complementary subspace, which depends on the phase $ \varphi $ through the longitudinal deformation parameter $ \lambda_\parallel(\varphi) $. It is precisely this second contribution which endows the probe with a linearly increasing sensitivity at large $ N $. It seems intuitively clear that singling out the relevant information contained in the subspace $ J^2 < N/2 $ requires a collective estimator, such as $ \hat{J}^2 $ itself. Note that such coherent manipulations may be demanding to implement, or even unavailable in case the probes are transmitted to $ N $ remote stations during the process.

Alternatively, one could ask about the performance of a collection of accessible \textit{separable} measurements, such as $ \hat{\sigma}_x^{\otimes N} $, implemented locally on each probe and supplemented by classical communication in the data analysis stage \cite{PhysRevLett.96.010401}. The corresponding phase sensitivity can also be computed analytically (see Appendix~\ref{app:parallel}), and quite remarkably it turns out to {\it coincide} with that of $ \hat{\sigma}_x $ in the $ N $-round setting (i.e., the dashed line in Fig.~\ref{fig2}). This behaviour is generic and does not depend on the specific sub-optimal observable measured on each probe. That is, although the parallel-entangled setting can, in principle, outperform the sequential one asymptotically \cite{demkowicz2014using}, we find that they may become metrologically equivalent when the probe readout at the output is limited to measuring local observables: This  restriction {\it de facto} banishes the asymptotic linear scaling of the precision.

It is important to remark that we assumed a particular probe preparation (GHZ states). Therefore, our observations should not be understood as a general `no-go' result, advocating against parallel-entangled estimation strategies in the presence of noise. The general question of whether the gap between sequential and parallel-entangled settings \cite{demkowicz2014using} persists when optimal input states and more general separable measurements are considered is definitely worthy of further investigation, although it lies beyond the scope of this paper.

\subsection{Sequential setting with passive ancilla}

Finally, as an example of the usefulness of entanglement in a practical sequential scenario, let us supplement the probe with a passive two-level ancilla. Specifically, we can prepare the probe-ancilla pair in a Bell state $ \ket{\Psi_\pm} \equiv \frac{1}{\sqrt{2}}\big(\ket{00}\pm\ket{11}\big) $ (which has the same $\ell_1$--norm of coherence as the single-qubit $\ket{+}$ and $N$-qubit GHZ states) and apply the noisy channel only on the first qubit, yielding $ \hat{\varrho}_N = (\Lambda_\varphi\otimes\mathbbm{1})^N\ket{\Psi_\pm}\bra{\Psi_\pm} $.

We find that, although interrogating probe and ancilla by a separable measurement like $ \hat{\sigma}_x\otimes\hat{\sigma}_x $ reduces once more to the same sensitivity as the single-qubit unassisted scenario (dashed curve in Fig.~\ref{fig2}), performing instead a non-separable (yet manageable) measurement such as $ \hat{O} = \ket{\Psi_+}\bra{\Psi_+}-\ket{\Psi_-}\bra{\Psi_-} $ does provide a sizeable increase in phase sensitivity (see Appendix~\ref{app:passive}).

\section{Discussion}\label{sec:d}

In general, one may conclude that using independent probes in a sequential scheme, possibly supplemented by correlated passive ancillas, offers a \textit{practical advantage} in noisy parameter estimation, in spite of the potential superiority of parallel-entangled strategies \cite{demkowicz2014using}. As we  illustrated, acquiring partial information about the geometry of the parameter-imprinting process allows one to optimize the estimation protocol at the single-probe level, by simply adjusting the sampling time or number of rounds. Such a sequential estimation protocol relies on the initial amount of ``unspeakable'' coherence \cite{Marvian2014,Marvian2016}, which is a genuinely quantum feature \cite{Baumgratz2014}, and is here confirmed as the key resource for estimating parameters encoded in incoherent operations, which include all phase-covariant channels. However, the estimation performance only scales linearly or ``classically'' in the probe size, whereby scaling up the probe size is intended as repeating the optimized sequential procedure $M \gg 1$ times using independent probe qubits, all initialized in a maximally coherent state. Nonetheless, at the single-probe level, the sensitivity does scale quadratically in the number of rounds $N$, provided $ N $ is well below $N_\text{opt}$.

Notably, in the technologically relevant limit of $ \varphi \ll 1 $ (e.g. magnetometry in a very weak magnetic field), the optimal number of rounds $ N_\text{opt} $ stays fairly large even for relatively low $ \kappa $ (see Appendix~\ref{app:kraus}), which translates into a very uncertain phase generator. As a result, a quadratic-like scaling of the precision for each individual probe can be maintained up to many iterations, although definitely not asymptotically. An interesting next step could be to extend our analysis to {\it multiparameter metrology}, e.g.~considering the simultaneous estimation of the phase $\varphi$ and the noise parameter $\kappa$, or the actual generator $\hat{H}_\mathbf{n}$ \cite{Datta2016}.

To conclude, let us remark that, in general, the comparison between different metrological settings is a particularly tricky subject, since all the \textit{resources} must be identified and properly accounted for \cite{giovannetti2011advances,Marvian2016,Kok,del2015resource}. For instance, in spite of the formal equivalence of sequential and parallel settings in absence of noise, promoting the `number of rounds' to the status of resource, at the same level as the `number of probes' is probably not fair, since the actual costs of preparation, control and measurement of an additional quantum probe are not comparable to the costs of increasing the sampling time for one additional round. It is surely worthwhile to put different metrological settings under a unified set-up, also including feedback control protocols, so as to carry out an objective bookkeeping of the associated costs. This will be the subject of future work.

\begin{acknowledgments}

We acknowledge the  European Research Council (ERC StG GQCOP, Grant No.~637352) and the Royal Society (Grant No.~IE150570) for financial support. We thank A. Datta,
R. Demkowicz-Dobrza{\'n}ski, A. Farace, C. Gogolin, M. Gu{\c{t}}\u{a}, L. Maccone, M. Mehboudi, K. Macieszczak,  K. Modi, and M. Oszmaniec for useful discussions, and especially J. Ko\l{}ody\'{n}ski for helping us improve the manuscript with his feedback.
\end{acknowledgments}

\onecolumngrid
\appendix
\setcounter{equation}{0}

\section{Lower bound for the QFI of a unital phase-covariant channel}\label{app:bound}

Below we will give further details about the tight-fitting lower bound $ f_N(\varphi) $ to the QFI $ F_N(\varphi) $. As stated in the main text, $\max_{\hat{\varrho}}\llangle\hat{\varrho}\vert \big(\partial_\varphi\mathcal{K}^N\big)^\dagger \partial_\varphi\mathcal{K}^N \vert\hat{\varrho}\rrangle \leq F_N(\varphi)$ \cite{alipour2014quantum,yousefjani2016framework} does hold in general although it is not necessarily a tight bound. If the maximization is not restricted to vectorized physical states $ \vert\hat{\varrho}\rrangle $, but extended to all normalized four-dimensional vectors in Liouville space, we would end up calculating the operator norm $ \vert\vert \big(\partial_\varphi\mathcal{K}^N\big)^\dagger \partial_\varphi\mathcal{K}^N \vert\vert \geq \max_{\hat{\varrho}}\llangle\hat{\varrho}\vert \big(\partial_\varphi\mathcal{K}^N\big)^\dagger \partial_\varphi\mathcal{K}^N \vert\hat{\varrho}\rrangle $. This equals the largest eigenvalue of the enclosed matrix. In particular, for the family of channels considered in the main text and represented by $ \mathcal{K}(\varphi) $, the eigenvalues of $ \big(\partial_\varphi\mathcal{K}^N\big)^\dagger \partial_\varphi\mathcal{K}^N $ are $ \eta_1 = 0 $, $ \eta_2 = N^2 \lambda_\parallel^{2N-2} (\partial_\varphi{\lambda}_\parallel)^2 $, and $ \eta_{3,4} = N^2 \lambda_\perp^{2N-2} [(\partial_\varphi\lambda_\perp)^2 + \lambda_\perp^2 (\partial_\varphi g)^2] $, which is doubly degenerate.

Now, considering the expression given in the main text for the QFI of any phase-covariant unital channel and recalling that $ 0 < \lambda_\perp < 1 $ one can readily see that $ \eta_{3,4}\equiv f_N(\varphi) \leq F_N(\varphi) $. Furthermore, whenever $ \eta_2 < \eta_{3,4} $ one could elegantly lower-bound the channel QFI as $ \vert\vert \big(\partial_\varphi\mathcal{K}^N\big)^\dagger \partial_\varphi\mathcal{K}^N \vert\vert = f_{N}(\varphi) \leq F_N(\varphi) $. This is the case, for instance, in the example considered in the main text, based on the von Mises-Fisher distribution. In the most general case, $ f_N(\varphi) $ still remains a close-fitting lower bound to the QFI for the whole class of channels represented by $ \mathcal{K}(\varphi) $.

\section{Operator-sum representation of the noisy phase-imprinting channel}\label{app:kraus}

\begin{figure}[b]
\centering
\includegraphics[width=0.9\linewidth]{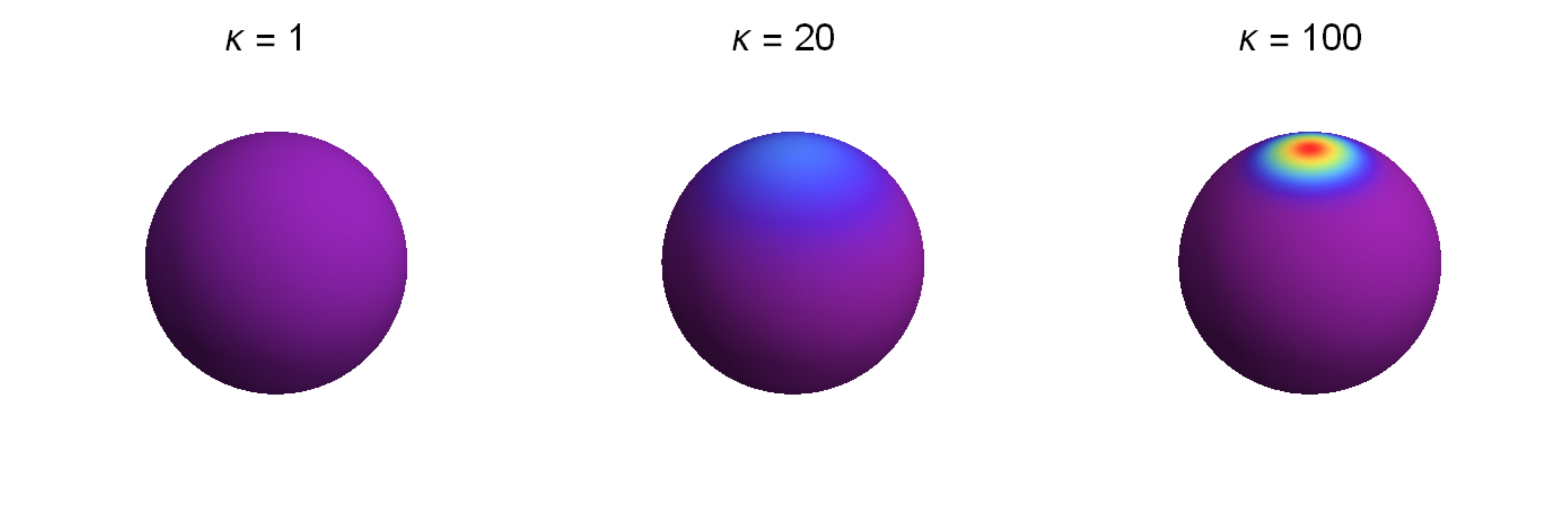}
\caption{(Color online) Visualization of the vMF distribution for varying $\kappa$. The probability density runs from indigo at its lowest to red at its highest. Note that the setting $ \kappa = 1 $, used to generate the illustrations of the main text, actually corresponds to a very broad distribution for the random generator $ \hat{H}_\mathbf{n} $. All the quantities plotted are dimensionless.
}
\label{fig:distspheres}
\end{figure}

In this section we shall give explicit expressions for a set of Kraus operators $ \{ \hat{K}_i \} $ realizing the noisy phase-imprinting channel $ \Lambda_\varphi\,\hat{\varrho} = \sum_i \hat{K_i}\hat{\varrho}\hat{K_i}^\dagger $ in Eq.~(1) of the main text, when the generic distribution $ p(\theta,\phi) $ for the random generator $ \hat{H}_\mathbf{n} $ is chosen to be the von Mises-Fisher distribution (vMF) distribution $ p_\kappa(\theta) $. Recall from the main text that $ p_\kappa(\theta) $ reads
\begin{equation}
p_\kappa(\theta)=\frac{\kappa e^{\kappa  \cos{\theta}}}{4 \pi \sinh \kappa},
\end{equation}
where the concentration parameter $\kappa  \geq 0$ gives an idea of how narrow the distribution of $ \mathbf{n} $ is around the $ z $--axis. For $\kappa\rightarrow 0$, the vMF distribution is uniform on the Bloch sphere, while it increases in concentration for greater $\kappa$. This is illustrated in  Fig.~\ref{fig:distspheres}.

The Kraus operators $ \{\hat{K}_i\} $ may be readily obtained from the eigenvalues and eigenvectors of the corresponding Choi matrix of the map $ \mathcal{C}_\Lambda $. This is the matrix resulting from $ \left(\Lambda_\varphi\otimes\mathbbm{1}\right)\ket{\Psi_+}\bra{\Psi_+} $, where $ \ket{\Psi_+} $ is the two-qubit Bell state $ \ket{\Psi_+} \equiv \frac{1}{\sqrt{2}} (\ket{00} + \ket{11}) $ \cite{breuer2002theory}. After a tedious but otherwise straightforward calculation one finds that the only non-zero matrix elements of the corresponding Kraus operators $ K_{i,kl} \equiv \bra{k}\hat{K}_i\ket{l} $ (with $ i\in\{1,\cdots,4\} $ and $ k,l\in\{ 1,0 \} $) are
\begin{equation}
\begin{split}
	K_{1,10} &= K_{2,01} = \frac{\sqrt{2}\sin{\varphi}}{\kappa}\sqrt{\kappa\coth{\kappa}-1}, \\
	K_{3,11} &= \frac{1}{2\sqrt{2}\kappa}\left(\sqrt{2\mathcal{A}+\mathcal{B}} + \sqrt{2\mathcal{A}-\mathcal{B}} \right), \qquad
	K_{4,11} = \frac{1}{2\sqrt{2}\kappa}\left(\sqrt{2\mathcal{A}+\mathcal{B}} - \sqrt{2\mathcal{A}-\mathcal{B}} \right), \\
	K_{3,00} &= \frac{\mathcal{C}}{\sqrt{2}}K_{3,11},\qquad K_{4,00} = -\frac{\mathcal{C}}{\sqrt{2}}K_{4,11},
	\label{eq:kraus_fmf}
\end{split}
\end{equation}
where $\mathcal{A}$ ($\in \mathbbm{R}$), $\mathcal{B}$ ($\in \mathbbm{R}$) and $\mathcal{C}$ ($\in \mathbbm{C}$) are the following functions of $\kappa$ and $\varphi$,
\begin{equation}
\begin{split}
	\mathcal{A}&=1+\kappa^2-\cos{2\varphi}-2\kappa\sin^2\varphi\,\coth{\kappa}, \\
	\mathcal{B}&=\sqrt{2\kappa^2(\cosh{2\kappa}-2\kappa^2-1)\csch^2{\kappa}\,\sin^2{2\varphi}}, \\
	\mathcal{C}&=\frac{\sqrt{\kappa^2\sin^2{2\varphi}\,(1+2\kappa^2-\cosh{2\kappa})+2[(1+\kappa^2-\cos{2\varphi})\sinh{\kappa}-2\kappa\cosh{\kappa}\sin^2{\varphi}]^2}}{\kappa[\kappa\sinh{(\kappa+2i\varphi)}-\cosh{(\kappa+2i\varphi)}+\cosh{\kappa}]-2\sin^2{\varphi}\sinh{\kappa}}.
	\label{eq:kraus_abc}
\end{split}
\end{equation}
This operator-sum representation then allows simple calculation of $\Lambda_{\varphi}\hat{\varrho}$ by bypassing the difficult integration that defines the channel.

The Liouville representation $ \mathcal{K}(\varphi) $ of the channel $ \Lambda_\varphi $ takes the form
\begin{equation}
\mathcal{K}(\varphi)=\left(\begin{array}{cccc}
\vert K_{3,00} \vert^2 + \vert K_{4,00} \vert^2 & 0 & 0 & K^2_{2,01} \\
0 & K_{3,00} K_{3,11} + K_{4,00} K_{4,11} & 0 & 0  \\
0 & 0 & K^{*}_{3,00} K_{3,11} + K^{*}_{4,11} K_{4,11} & 0 \\
K^{2}_{1,10} & 0 & 0 & K^{2}_{3,11} + K^{2}_{4,11}
\end{array}\right),
\label{eq:liouville_gen}
\end{equation}
in terms of the operator-sum representation of the channel. In particular, we have that
\begin{equation}
\lambda_\parallel = 1-2 K^{2}_{2,01}, \qquad \qquad
\lambda_\perp =  \vert S \vert, \qquad \text{and} \qquad g = -\mbox{arg}(S),
\label{lambdas_fmf}
\end{equation}
where $S= \frac{\sqrt{4\mathcal{A}^2-\mathcal{B}^2}}{2\sqrt{2}\kappa^2} \mathcal{C}$ ($\in \mathbbm{C}$) and $\mbox{arg}(S)$ denotes the complex argument of $S$.

\section{Phase sensitivity in the sequential setting}\label{app:sequential}

\subsection{Quantum Fisher information}

We will now provide a closed formula for the QFI $ F_N(\varphi) $ for a single two-level probe undergoing $ N $ sequential applications of $ \Lambda_\varphi $. We shall take the optimal `plus' state $\ket{+}=\frac{1}{\sqrt{2}}(\ket{0}+\ket{1})$ as the probe preparation $ \hat{\varrho}_0 $.

Recall from the main text that $ F_N(\varphi) $ of $ \hat{\varrho}_N $ simply writes as
\begin{equation}\label{eq:QFIgeneral}
F_{N}(\varphi)=4 \sum_{i,j} \frac{q_i}{(q_i+q_j)^2}\left\vert \bra{\psi_i}\partial_\varphi\,\hat{\varrho}_N\vert\psi_j\rangle \right\vert^2 ,
\end{equation}
where $q_i $ and $ \ket{\psi_i} $ are the eigenvalues and eigenvectors of $\hat{\varrho}_{N}$, and any terms for which $q_{i}+q_{j}=0$ are excluded from the sum. The output state of the probe $\hat{\varrho}_{N} = \Lambda^{N}_{\varphi}\ket{+}\bra{+}$ can be obtained by repeatedly applying the Kraus operators defined in Eqs.~\eqref{eq:kraus_fmf} and \eqref{eq:kraus_abc}. After a lengthy calculation, one may cast $ F_N(\varphi) $ in compact form as
\begin{equation}
F_{N}(\varphi)=N^2  \left| S\right| ^{2 N} \left| \frac{S'}{S}\right| ^2\frac{1- \left| S\right| ^{2 N}\sin^2(\nu-\mu)}{1- \left| S\right| ^{2 N}},
\label{eq:seqQFI}
\end{equation}
where the phases $ \mu $ and $ \nu $ are $ \mu \equiv \arg{S} $  and $ \nu\equiv\arg{S'} = \arg{\left(\partial_\varphi S\right)} $. As expected from the discussion in the main text, the QFI of Eq.~\eqref{eq:seqQFI} does grow as $N^2$ before a peak is reached at some optimal $ N_{\text{opt}} $, after which there is a noise-dependent exponential decay, asymptotically approaching zero in the limit of large $N$.

\subsection{Lower bound on the quantum Fisher information}

\begin{figure}
	\includegraphics[width=7.5cm]{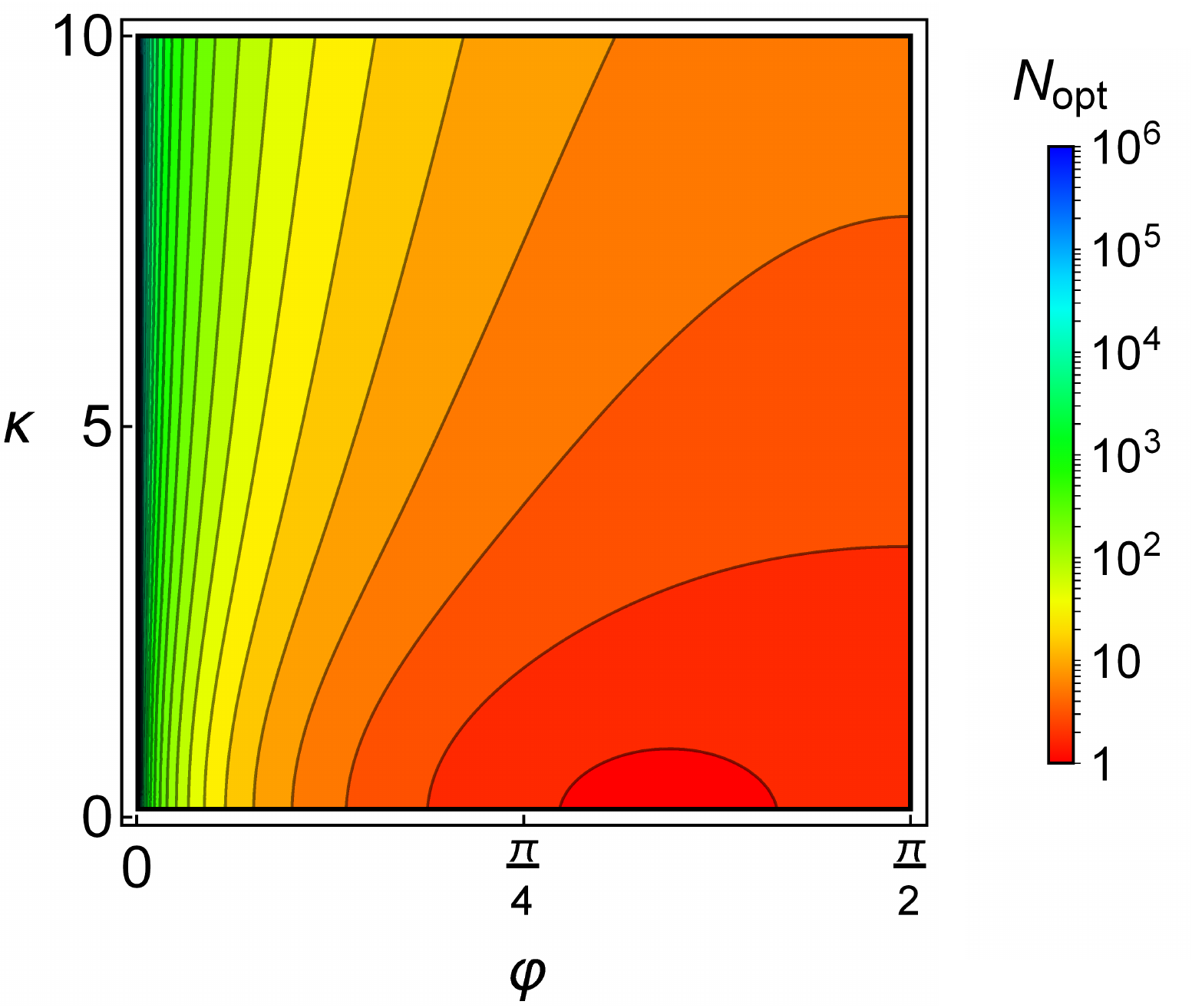}
	\caption{(Color online) Contour plot of the optimal number of iterations $ N_\text{opt} $ as obtained from the maximization of $ f_{N}(\varphi) $, as a function of the phase $ \varphi $ and the concentration parameter $ \kappa $. All the quantities plotted are dimensionless.}
\label{fig:contour}
\end{figure}

We will now comment on the expression of the lower bound $ f_{N}(\varphi) $ on the QFI for our particular noisy channel, in the sequential setting. In this particular case $ f_{N}(\varphi) = N^2\, \lambda_\perp^{2N-2} [(\partial_\varphi\lambda_\perp)^2 + \lambda_\perp^2 (\partial_\varphi g)^2] = N^2 \vert S \vert^{2N} \vert S'/S \vert^2 $. Note that $f_{N}(\varphi) $ is thus just the prefactor in the expression for $ F_N(\varphi) $ in of Eq.~\eqref{eq:seqQFI}, which essentially modulates the amplitude of the optimal phase sensitivity.

In Fig.~\ref{fig:contour}, we plot $ N_\text{opt} = -1/\log{\vert S \vert} $ as a function of the concentration parameter $ \kappa $ and the phase $ \varphi $. Note that for $ \varphi $ small enough, the optimal number of applications remains on the order of $ 10^3 $ even for very low $ \kappa $ (i.e. quasi-uniform distribution for the rotation axis $ \mathbf{n} $). As a result, the sequential setting may exhibit a super-classical scaling in the sensitivity, up to a significantly large number of rounds.

\subsection{Phase sensitivity of $ \hat{\sigma}_x $}

The maximum phase sensitivity, given by the QFI, may only be reached when an optimal estimator is measured on the output state of the probe. Recall that such optimal estimator must be diagonal in the eigenbasis of the SLD $ \hat{L} $, which reads \cite{barndorff2000fisher}
\begin{equation}\label{Eq:SLD}
\hat{L}=2\sum_{ij} \frac{\langle \psi_i \vert \partial_\varphi\hat{\varrho}_N \vert \psi_j \rangle }{q_i+q_j}\vert\psi_i\rangle\langle\psi_j\vert.
\end{equation}
We can instead calculate the phase sensitivity $ F^{\hat{O}} $ of some sub-optimal observable $ \hat{O} $ 
defined as in Eq.~(\ref{eq:cfi_general}).
Choosing $ \hat{\sigma}_x $ as our estimator yields
\begin{equation}\label{Eq:SeqCFI}
F^{\hat{\sigma}_x}=N^2  \left| S\right| ^{2 N} \left| \frac{S'}{S}\right| ^2 \frac{\cos ^2(\nu +(N-1) \mu )}{1-\left| S\right| ^{2 N} \cos ^2(N \mu )}.
\end{equation}

\begin{figure}[h!]
\centering
\includegraphics[width=0.7\linewidth]{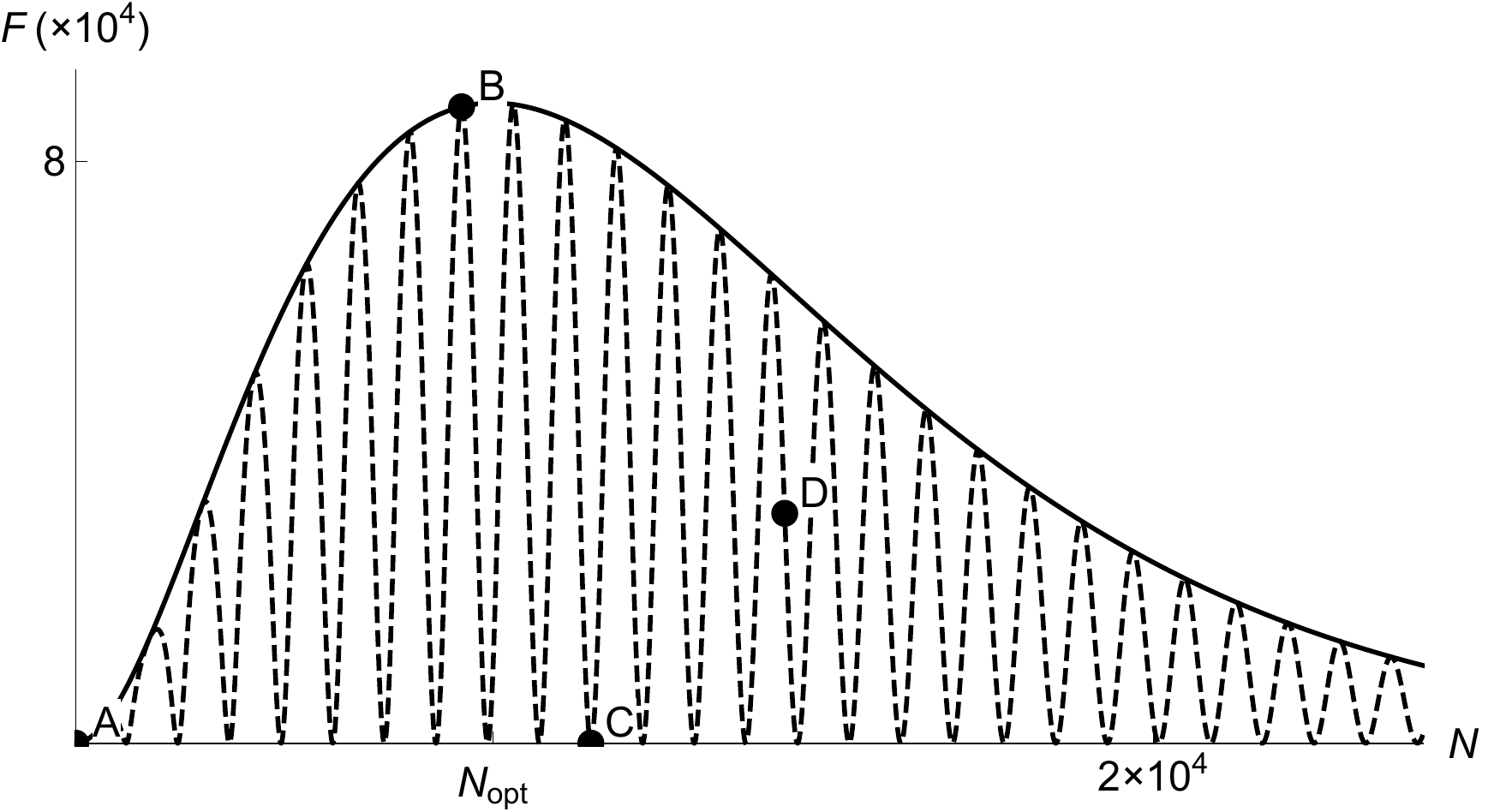}\\
\includegraphics[width=0.245\linewidth]{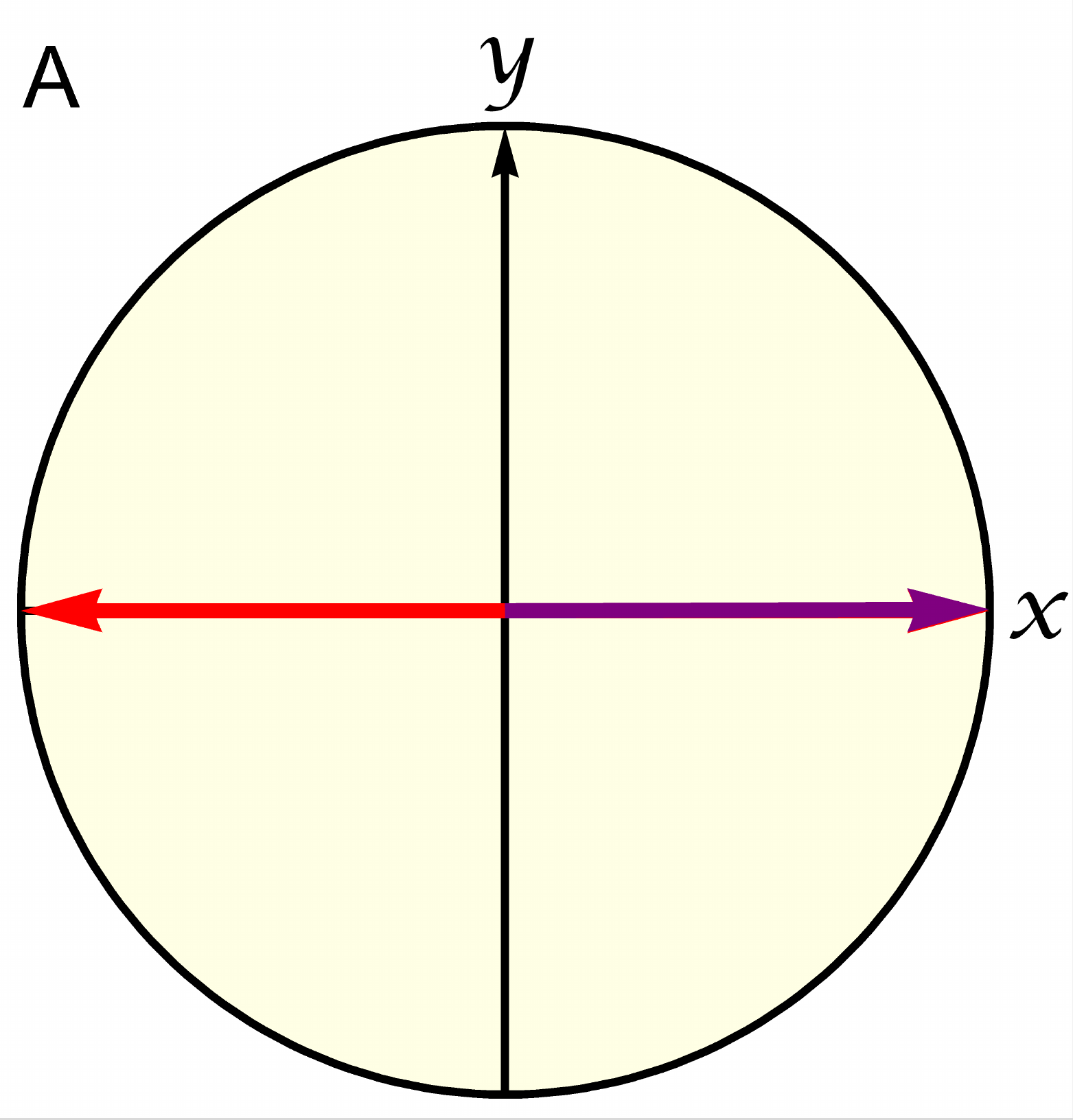}
\includegraphics[width=0.245\linewidth]{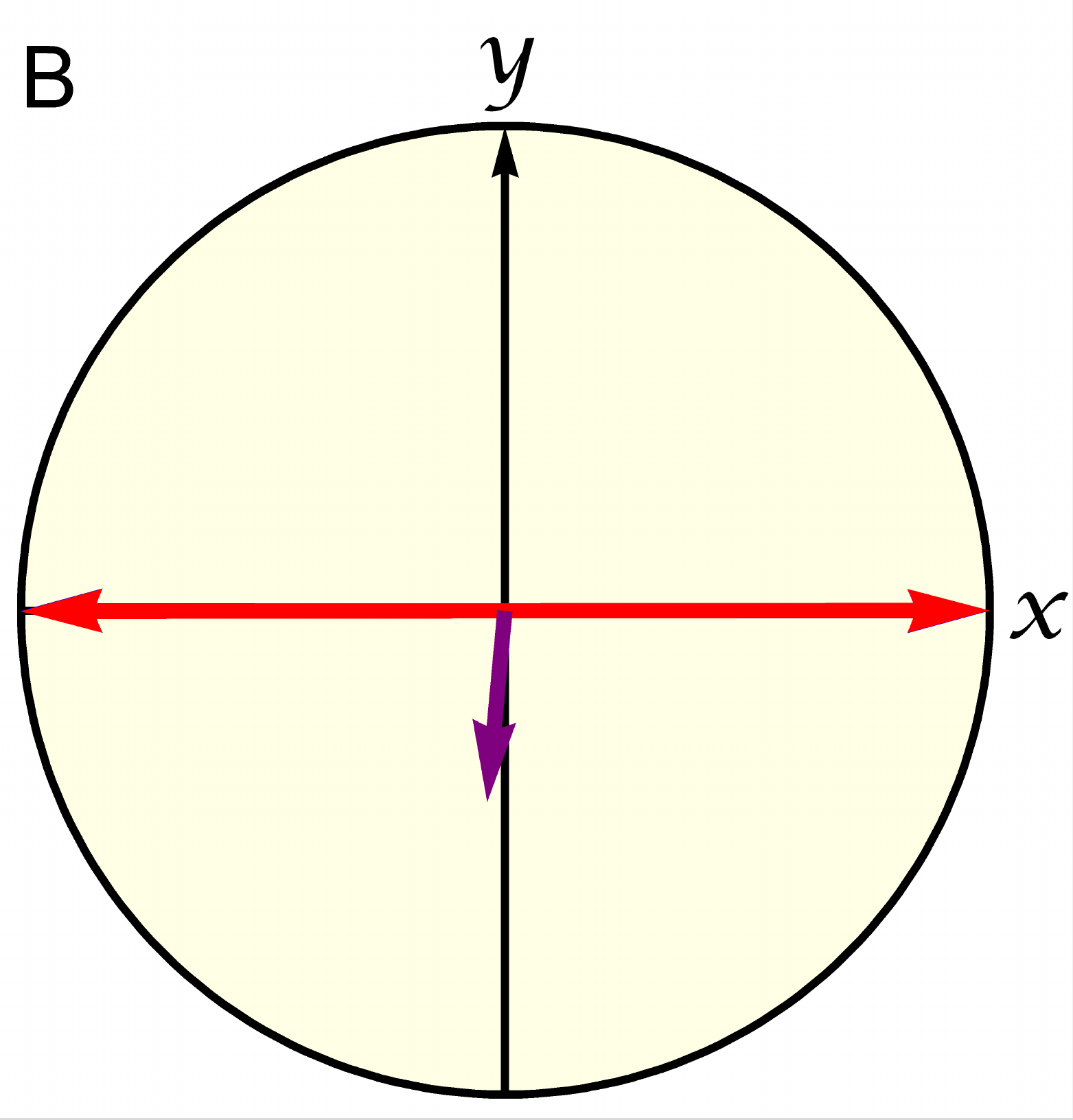}
\includegraphics[width=0.245\linewidth]{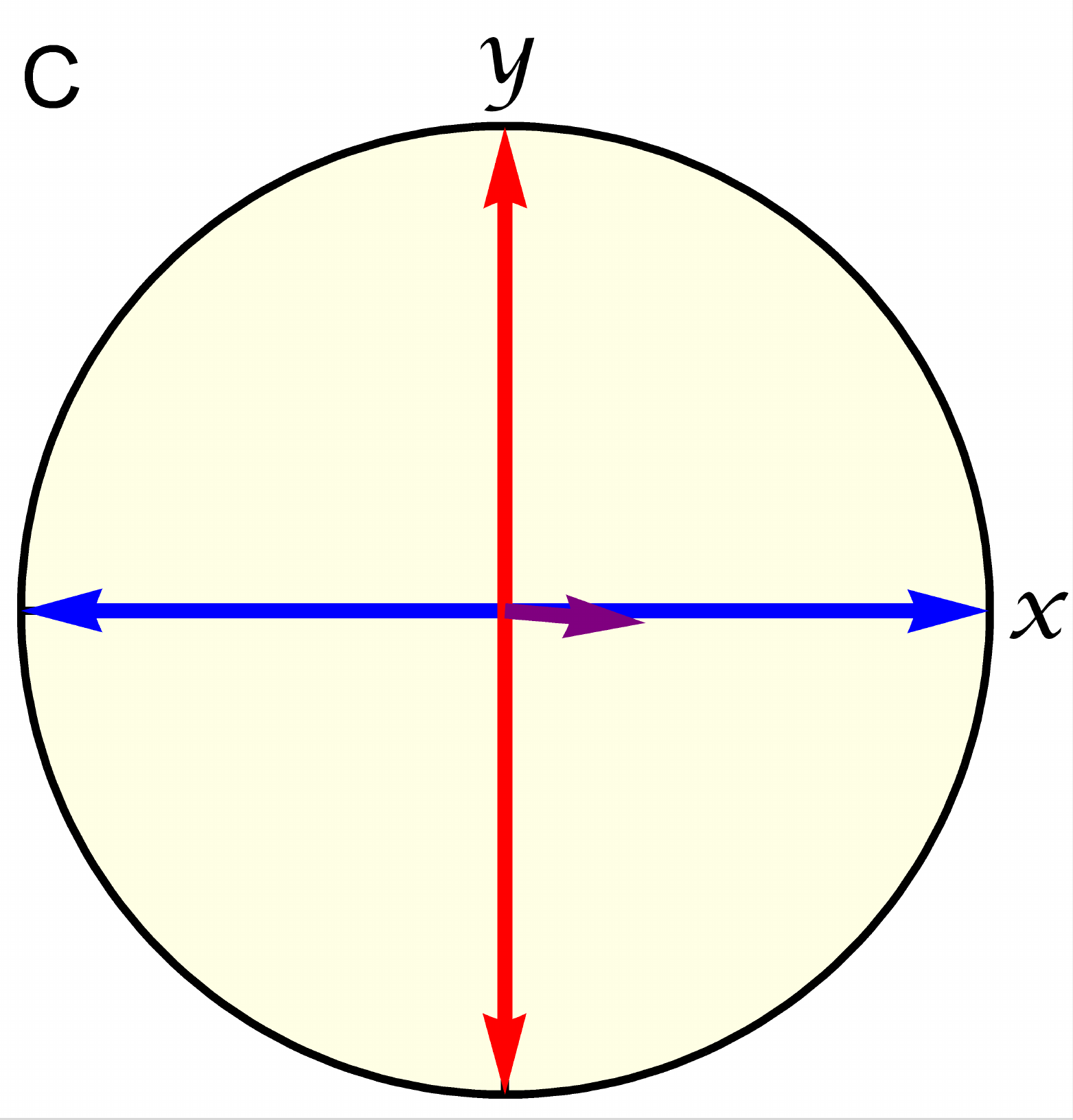}
\includegraphics[width=0.245\linewidth]{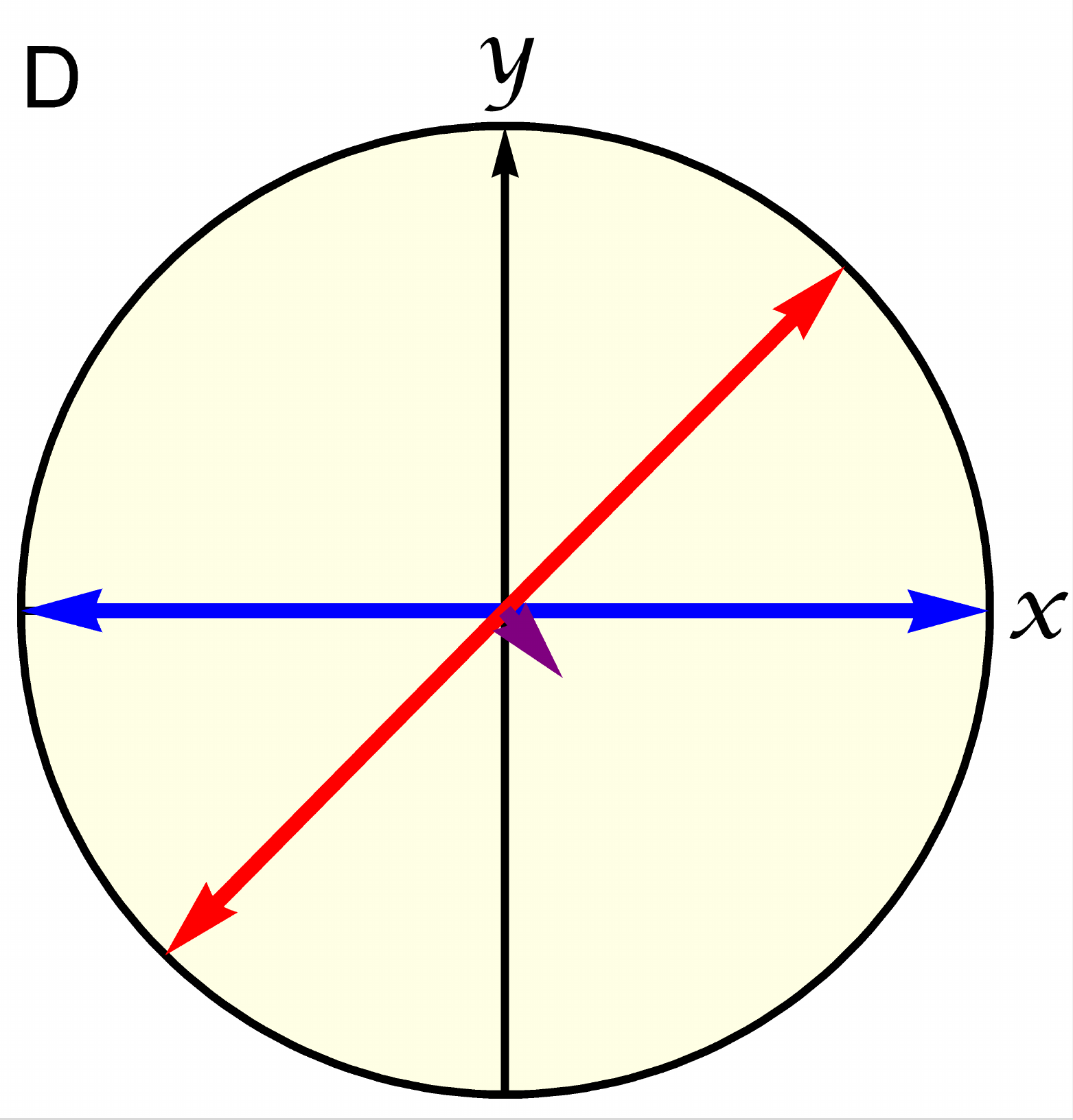}
\caption{(Color online) Top: QFI (solid line) and phase sensitivity of $\hat{\sigma}_{x}$ (dashed line) for $\varphi=0.01$ and $\kappa=0.5$. Representations of the evolved probe state (purple arrow), optimal measurement basis (red arrow), and sub-optimal measurement basis (blue arrow), as Bloch vectors on the equatorial plane are shown in the bottom panels. (A) Initially, the probe state and both measurements are aligned. (B) The phase sensitivity of $ \hat{\sigma}_{x} $ meets the QFI when the measurement bases realign. (C) The phase sensitivity of $\hat{\sigma}_{x}$ vanishes whenever the measurement bases become perpendicular. (D) Otherwise, the phase sensitivity of $\hat{\sigma}_{x}$ oscillates between zero and the QFI. Note that, as $ N $ grows, the optimal basis vectors become perpendicular to the probe state vector as they rotate. Note as well that the probe state vector gradually shortens due to the loss of purity. All the quantities plotted are dimensionless.}
\label{fig:blochA}
\end{figure}

The QFI and the phase sensitivity $F^{\hat{\sigma}_x}$ are depicted in Fig.~\ref{fig:blochA} for particular values of $\varphi$ and $\kappa$. It can be seen that $F_{N}(\varphi)$ displays the aforementioned quadratic behaviour followed by an exponential tail-off, whereas $F^{\hat{\sigma}_x}$ oscillates between zero and $F_{N}(\varphi)$. This curious behaviour can be understood by visualizing the evolved probe state $\hat{\varrho}_{N}$ and the measurement eigenbases of both the SLD $\hat{L}$ and the sub-optimal estimator $\hat{\sigma}_{x}$ on the equatorial plane of the Bloch sphere (see panels A-D in Fig.~\ref{fig:blochA}).

Here, the initial probe state $\hat{\varrho}_{0}=\ket{+}\bra{+}$ begins on the $x$--axis at the surface of the sphere. As stated in the main text, the Bloch vector of the evolved state $\hat{\varrho}_{N}$ \textit{inspirals} towards the normalized identity. This is a result of the rotation around the $z$--axis due to the parameter-encoding unitary, and the loss of purity due to the noise. The optimal measurement basis begins on the $x$--axis, parallel to the probe state, and  rotates with $N$. Meanwhile, the fixed measurement basis lies on the $ x $--axis, so that it periodically coincides with the optimal one: when they are parallel, $ F^{\hat{\sigma}_x} = F_{N}(\varphi) $, whereas when they become perpendicular, $ F^{\hat{\sigma}_x} = 0 $. The frequency of these oscillations is given approximately by $ |\mu|/\pi $.

\section{Phase sensitivity in the sequential setting with a passive ancilla}\label{app:passive}

In this section we will give further details about the performance of the sequential estimation setting when the probe is supplemented with a passive two-level ancilla. Recall from the main text that, in this case, we prepare the two qubits in a Bell state $ \ket{\Psi_\pm} = \frac{1}{\sqrt{2}}(\ket{00}\pm\ket{11}) $ and proceed to apply sequentially $ N $ times the phase-imprinting channel as $ \hat{\varrho}_N = (\Lambda_\varphi\otimes\mathbbm{1})^N\ket{\Psi_\pm}\bra{\Psi_\pm} $.

The resulting `evolved' two-qubit state has maximally mixed marginals at any $ N $. The corresponding QFI $ F_N^{(\text{anc})}(\varphi) $ may be readily evaluated to
\begin{equation}
F_N^{(\text{anc})}(\varphi) = \frac14 N^2\left(\mathcal{D}_+ + \mathcal{D}_- + \frac{8\lambda_\perp^{2N}(\partial_\varphi g)^2}{1+\lambda_\parallel^N} + \frac{2\lambda_\parallel^{2(N-1)}(\partial_\varphi \lambda_\parallel)^2}{1-\lambda_\parallel^N} \right),
\label{eq:qfi_seq_anc}
\end{equation}
where
\begin{equation}
\mathcal{D}_\pm\equiv\frac{\vert 2\lambda_\perp^{N-1}(\partial_\varphi\lambda_\perp)\pm\lambda_\parallel^{N-1}(\partial_\varphi\lambda_\parallel) \vert^2}{1\pm 2\lambda_\perp^N+\lambda_\parallel^N}.
\label{eq:qfi_pieces}
\end{equation}

\begin{figure}[h!]
\centering
\includegraphics[width=0.5\linewidth]{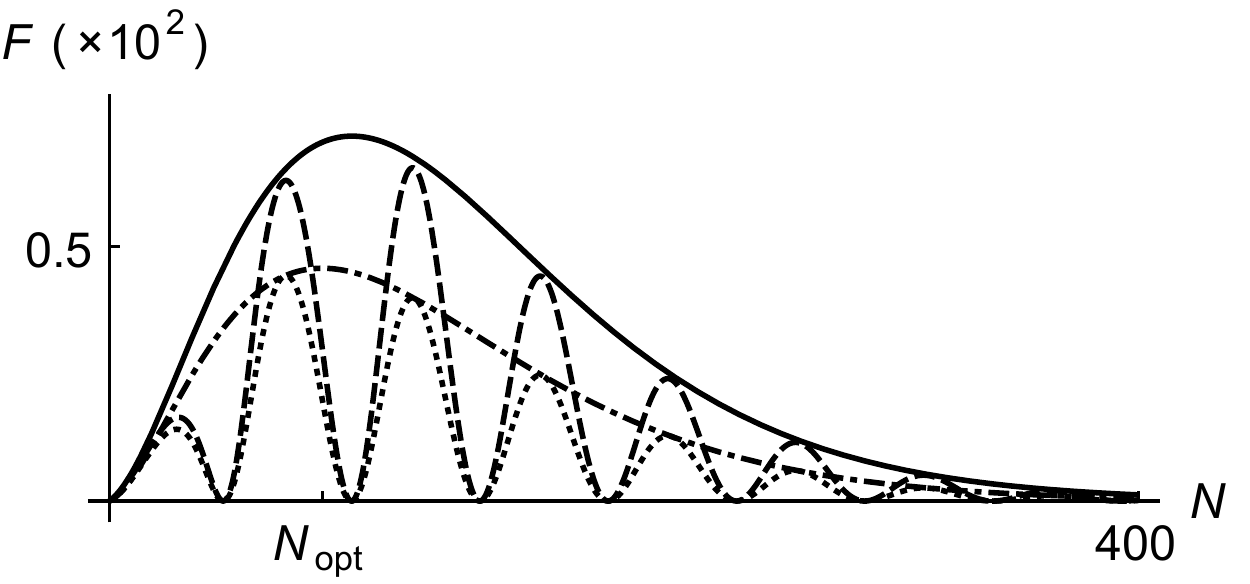}
\caption{QFI for the sequential setting when the probe is supplemented with a passive ancilla (solid) as a function of the number of applications of the channel $ N $. The QFI for a single two-level probe (without ancilla) is included for comparison (dot-dashed). The phase sensitivity of $ \hat{\sigma}_x\otimes\hat{\sigma}_x $ (dotted) and $ \hat{O} \equiv \ket{\Psi_+}\bra{\Psi_+} - \ket{\Psi_-}\bra{\Psi_-} $ (dashed) have also been plotted. The vMF distribution is assumed for the stochastic generator of the phase rotations, with $ \varphi = 0.1 $ and $ \kappa = 1 $. All the quantities plotted are dimensionless.}
\label{fig:seq_anc}
\end{figure}

Once again, we shall particularize our results choosing the vMF distribution for the stochastic generator (see Fig.~\ref{fig:seq_anc}). In the first place, note that $ F_N^{(\text{anc})}(\varphi) $ notably outperforms the sensitivity of a single probe. In particular, when a separable estimator such as $ \hat{\sigma}_x\otimes\hat{\sigma}_x $ is considered, the phase sensitivity of the probe-ancilla pair remains upper-bounded by the QFI of a single probe. However, as we can see, when a Bell measurement is performed \textit{jointly} on the probe and the ancilla, the ensuing sensitivity can come much closer to the ultimate limit set by $ F_N^{(\text{anc})}(\varphi) $.

\section{Phase sensitivity in the parallel setting}\label{app:parallel}

\subsection{Quantum Fisher information}

We will now compute the QFI in the parallel setting for a maximally entangled Greenberger-Horne-Zeilinger (GHZ) state of the $ N $ probes, $\ket{\mathrm{GHZ}}\equiv\frac{1}{\sqrt{2}}\left(\ket{0}^{\otimes N}+\ket{1}^{\otimes N}\right)$. The output state $\hat{\varrho}_{N} =\Lambda_{\varphi}^{\otimes N}\ket{\mathrm{GHZ}}\bra{\mathrm{GHZ}}$ is a type of \textit{X--state}, that is, a state represented by a $ 2^N\times 2^N $ matrix with non-zero elements $ \hat{\varrho}_{N,mn} = \bra{m}\hat{\varrho}_N\ket{n} $ (with $m,n \in \{0,1,\cdots ,2^{N}-1\}$) only along the main diagonal and in the extreme off-diagonal corners. In particular, we find
\begin{equation}
\begin{split}
 \rho_{N,mm} & = \alpha^{N-H(m)}(1-\alpha)^{H(m)}+\alpha^{H(m)}(1-\alpha)^{N-H(m)} \\
 \rho_{N,0 (2^N-1)}& = S^N \\
 \rho_{N,(2^N-1) 0}& = \bar{S}^{N}, \\
\end{split}
\label{eq:parallel_state}
\end{equation}
where $H(m)$ is the Hamming weight of $m$, i.e. the number of $1$'s in the binary representation of $m$. The over-bar denotes complex conjugation and $\alpha$ is given by
\begin{equation}
\alpha = \frac12(1-\lambda_\parallel) = \frac{2 (-1 + \kappa \coth\kappa) \sin^2\varphi}{\kappa^2}.
\end{equation}
Then, the QFI ($ F_N^\text{par} $) may be calculated using Eq.~\eqref{eq:QFIgeneral} and is given by
\begin{multline}
F_N^{\text{par}}=\frac{N^2}{8} \left(-\frac{
8\left| S\right| ^{2N} \left| \frac{S'}{S}\right|^2 \sin^2(\nu-\mu)}{(1 - \alpha)^N + \alpha^
N}
 + \frac{(2\left| S\right| ^N \left| \frac{S'}{S}\right| \sin(\nu-\mu)+
4 ((1 - \alpha)^{N-1} \alpha - \alpha^
N) \cot\varphi)^2}{(1 - \alpha)^N + \alpha^N - \lvert S\rvert^
N}\right.\\
\left.
 + \frac{(2\left| S\right| ^N \left| \frac{S'}{S}\right| \sin(\nu-\mu) -
4 ((1 - \alpha)^{N-1} \alpha - \alpha^N) \cot\varphi)^2}{(1 - \alpha)^N + \alpha^N + \lvert S\rvert^N}\right)\\
 + 2 \sum_{k=1}^{N-1}  \left( \begin{array}{c} N \\
 k \end{array} \right)
 \frac{\left(\frac{(k + N (\alpha-1)) \left(\frac{1}{\alpha}-1\right)^k \alpha^
N}{\alpha-1} + (1 - \alpha)^{N-1-k} \alpha^
k (k - N \alpha)\right)^2 \cot^2\varphi}{\left(\frac{1}{\alpha}-1\right)^
k \alpha^N + (1 - \alpha)^{N-k} \alpha^k}.
\label{eq:parallel_qfi}
\end{multline}

The contribution of the first term in Eq.~\eqref{eq:parallel_qfi} to the total sensitivity is qualitatively similar to the QFI in the sequential setting, i.e. it scales quadratically for low $ N $ and, after peaking, it decays exponentially to zero. However, the second term, contributes with an classical-like increase in $ N $, which ultimately yields a linear asymptotic scaling for the overall QFI. As stated in the main text, while the first contribution relates to the outermost ``corners'' of the density operator (i.e. the subspace spanned by $ \ket{00\cdots 0} $ and $ \ket{11\cdots 1} $, with total angular momentum $ J = 1/2 $), the second one is determined by all the matrix elements along the diagonal of the state, such that $ J < 1/2 $. In particular, accessing all the relevant information giving rise to the classical-like asymptotic scaling of the sensitivity thus requires one to perform a non-separable measurement capable of telling the subspace $ J = 1/2 $ from its complementary $ J < 1/2 $, such as a measurement of the total angular momentum $ \hat{J}^2 $.

\subsection{Phase sensitivity of $ \hat{\sigma}_x^{\otimes N} $}

Analogously to the sequential setting, one can consider the sensitivity of a sub-optimal estimator. In particular, we can compute the phase sensitivity of the separable observable $ \hat{O} = \hat{\sigma}_{x}^{\otimes N} $ by resorting to Eq.~\eqref{eq:cfi_general} and \eqref{eq:parallel_state}. As pointed out in the main text this yields exactly the same formula of Eq.~\eqref{Eq:SeqCFI}, thus implying that the parallel-entangled and unentangled sequential settings are metrologically equivalent so long as the estimation is constrained to separable measurements.

Note furthermore that, in both sequential and parallel settings, one could consider alternative observables to $\hat{\sigma}_{x}$ with eigenbases on the equatorial plane of the Bloch sphere, such as $\hat{\sigma}_{y}$ or even some suboptimal yet phase-dependent observable. Generically, the oscillatory behavior will remain, but the periodicity and locations of the maxima will change depending on the chosen observable.


\twocolumngrid

%

\begin{thebibliography}{49}
\expandafter\ifx\csname natexlab\endcsname\relax\def\natexlab#1{#1}\fi
\expandafter\ifx\csname bibnamefont\endcsname\relax
  \def\bibnamefont#1{#1}\fi
\expandafter\ifx\csname bibfnamefont\endcsname\relax
  \def\bibfnamefont#1{#1}\fi
\expandafter\ifx\csname citenamefont\endcsname\relax
  \def\citenamefont#1{#1}\fi
\expandafter\ifx\csname url\endcsname\relax
  \def\url#1{\texttt{#1}}\fi
\expandafter\ifx\csname urlprefix\endcsname\relax\def\urlprefix{URL }\fi
\providecommand{\bibinfo}[2]{#2}
\providecommand{\eprint}[2][]{\url{#2}}

\bibitem[{\citenamefont{{B. P. Abbott {\it et al.}}}(2016)}]{LIGO}
\bibinfo{author}{\bibnamefont{{B. P. Abbott {\it et al.}}}}
  (\bibinfo{collaboration}{LIGO Scientific Collaboration and Virgo
  Collaboration}), \bibinfo{journal}{Phys. Rev. Lett.}
  \textbf{\bibinfo{volume}{116}}, \bibinfo{pages}{061102}
  (\bibinfo{year}{2016}).

\bibitem[{\citenamefont{Giovannetti et~al.}(2004)\citenamefont{Giovannetti,
  Lloyd, and Maccone}}]{giovannetti2004quantum}
\bibinfo{author}{\bibfnamefont{V.}~\bibnamefont{Giovannetti}},
  \bibinfo{author}{\bibfnamefont{S.}~\bibnamefont{Lloyd}}, \bibnamefont{and}
  \bibinfo{author}{\bibfnamefont{L.}~\bibnamefont{Maccone}},
  \bibinfo{journal}{Science} \textbf{\bibinfo{volume}{306}},
  \bibinfo{pages}{1330} (\bibinfo{year}{2004}).

\bibitem[{\citenamefont{Giovannetti et~al.}(2006)\citenamefont{Giovannetti,
  Lloyd, and Maccone}}]{PhysRevLett.96.010401}
\bibinfo{author}{\bibfnamefont{V.}~\bibnamefont{Giovannetti}},
  \bibinfo{author}{\bibfnamefont{S.}~\bibnamefont{Lloyd}}, \bibnamefont{and}
  \bibinfo{author}{\bibfnamefont{L.}~\bibnamefont{Maccone}},
  \bibinfo{journal}{Phys. Rev. Lett.} \textbf{\bibinfo{volume}{96}},
  \bibinfo{pages}{010401} (\bibinfo{year}{2006}).

\bibitem[{\citenamefont{Giovannetti et~al.}(2011)\citenamefont{Giovannetti,
  Lloyd, and Maccone}}]{giovannetti2011advances}
\bibinfo{author}{\bibfnamefont{V.}~\bibnamefont{Giovannetti}},
  \bibinfo{author}{\bibfnamefont{S.}~\bibnamefont{Lloyd}}, \bibnamefont{and}
  \bibinfo{author}{\bibfnamefont{L.}~\bibnamefont{Maccone}},
  \bibinfo{journal}{Nat. Phot.} \textbf{\bibinfo{volume}{5}},
  \bibinfo{pages}{222} (\bibinfo{year}{2011}).

\bibitem[{\citenamefont{Demkowicz-Dobrza{\'n}ski
  et~al.}(2012)\citenamefont{Demkowicz-Dobrza{\'n}ski, Ko{\l}ody{\'n}ski, and
  Gu{\c{t}}\u{a}}}]{demkowicz2012elusive}
\bibinfo{author}{\bibfnamefont{R.}~\bibnamefont{Demkowicz-Dobrza{\'n}ski}},
  \bibinfo{author}{\bibfnamefont{J.}~\bibnamefont{Ko{\l}ody{\'n}ski}},
  \bibnamefont{and}
  \bibinfo{author}{\bibfnamefont{M.}~\bibnamefont{Gu{\c{t}}\u{a}}},
  \bibinfo{journal}{Nat. Commun.} \textbf{\bibinfo{volume}{3}},
  \bibinfo{pages}{1063} (\bibinfo{year}{2012}).

\bibitem[{\citenamefont{Huelga et~al.}(1997)\citenamefont{Huelga, Macchiavello,
  Pellizzari, Ekert, Plenio, and Cirac}}]{PhysRevLett.79.3865}
\bibinfo{author}{\bibfnamefont{S.~F.} \bibnamefont{Huelga}},
  \bibinfo{author}{\bibfnamefont{C.}~\bibnamefont{Macchiavello}},
  \bibinfo{author}{\bibfnamefont{T.}~\bibnamefont{Pellizzari}},
  \bibinfo{author}{\bibfnamefont{A.~K.} \bibnamefont{Ekert}},
  \bibinfo{author}{\bibfnamefont{M.~B.} \bibnamefont{Plenio}},
  \bibnamefont{and} \bibinfo{author}{\bibfnamefont{J.~I.} \bibnamefont{Cirac}},
  \bibinfo{journal}{Phys. Rev. Lett.} \textbf{\bibinfo{volume}{79}},
  \bibinfo{pages}{3865} (\bibinfo{year}{1997}).

\bibitem[{\citenamefont{Escher et~al.}(2011)\citenamefont{Escher,
  de~Matos~Filho, and Davidovich}}]{escher2011noisy}
\bibinfo{author}{\bibfnamefont{B.}~\bibnamefont{Escher}},
  \bibinfo{author}{\bibfnamefont{R.}~\bibnamefont{de~Matos~Filho}},
  \bibnamefont{and}
  \bibinfo{author}{\bibfnamefont{L.}~\bibnamefont{Davidovich}},
  \bibinfo{journal}{Nat. Phys.} \textbf{\bibinfo{volume}{7}},
  \bibinfo{pages}{406} (\bibinfo{year}{2011}).

\bibitem[{\citenamefont{Matsuzaki et~al.}(2011)\citenamefont{Matsuzaki,
  Benjamin, and Fitzsimons}}]{matsuzaki2011magnetic}
\bibinfo{author}{\bibfnamefont{Y.}~\bibnamefont{Matsuzaki}},
  \bibinfo{author}{\bibfnamefont{S.~C.} \bibnamefont{Benjamin}},
  \bibnamefont{and}
  \bibinfo{author}{\bibfnamefont{J.}~\bibnamefont{Fitzsimons}},
  \bibinfo{journal}{Phys. Rev. A} \textbf{\bibinfo{volume}{84}},
  \bibinfo{pages}{012103} (\bibinfo{year}{2011}).

\bibitem[{\citenamefont{Chin et~al.}(2012)\citenamefont{Chin, Huelga, and
  Plenio}}]{PhysRevLett.109.233601}
\bibinfo{author}{\bibfnamefont{A.~W.} \bibnamefont{Chin}},
  \bibinfo{author}{\bibfnamefont{S.~F.} \bibnamefont{Huelga}},
  \bibnamefont{and} \bibinfo{author}{\bibfnamefont{M.~B.}
  \bibnamefont{Plenio}}, \bibinfo{journal}{Phys. Rev. Lett.}
  \textbf{\bibinfo{volume}{109}}, \bibinfo{pages}{233601}
  (\bibinfo{year}{2012}).

\bibitem[{\citenamefont{Macieszczak}(2015)}]{PhysRevA.92.010102}
\bibinfo{author}{\bibfnamefont{K.}~\bibnamefont{Macieszczak}},
  \bibinfo{journal}{Phys. Rev. A} \textbf{\bibinfo{volume}{92}},
  \bibinfo{pages}{010102} (\bibinfo{year}{2015}).

\bibitem[{\citenamefont{Smirne et~al.}(2016)\citenamefont{Smirne,
  {Ko\l{}ody\ifmmode \acute{n}\else \'{n}\fi{}ski}, Huelga, and
  {Demkowicz-Dobrza\ifmmode \acute{n}\else
  \'{n}\fi{}ski}}}]{smirne2015ultimate}
\bibinfo{author}{\bibfnamefont{A.}~\bibnamefont{Smirne}},
  \bibinfo{author}{\bibfnamefont{J.}~\bibnamefont{{Ko\l{}ody\ifmmode
  \acute{n}\else \'{n}\fi{}ski}}}, \bibinfo{author}{\bibfnamefont{S.~F.}
  \bibnamefont{Huelga}}, \bibnamefont{and}
  \bibinfo{author}{\bibfnamefont{R.}~\bibnamefont{{Demkowicz-Dobrza\ifmmode
  \acute{n}\else \'{n}\fi{}ski}}}, \bibinfo{journal}{Phys. Rev. Lett.}
  \textbf{\bibinfo{volume}{116}}, \bibinfo{pages}{120801}
  (\bibinfo{year}{2016}).

\bibitem[{\citenamefont{Tanaka et~al.}(2015)\citenamefont{Tanaka, Knott,
  Matsuzaki, Dooley, Yamaguchi, Munro, and Saito}}]{tanaka2015proposed}
\bibinfo{author}{\bibfnamefont{T.}~\bibnamefont{Tanaka}},
  \bibinfo{author}{\bibfnamefont{P.}~\bibnamefont{Knott}},
  \bibinfo{author}{\bibfnamefont{Y.}~\bibnamefont{Matsuzaki}},
  \bibinfo{author}{\bibfnamefont{S.}~\bibnamefont{Dooley}},
  \bibinfo{author}{\bibfnamefont{H.}~\bibnamefont{Yamaguchi}},
  \bibinfo{author}{\bibfnamefont{W.~J.} \bibnamefont{Munro}}, \bibnamefont{and}
  \bibinfo{author}{\bibfnamefont{S.}~\bibnamefont{Saito}},
  \bibinfo{journal}{Phys. Rev. Lett.} \textbf{\bibinfo{volume}{115}},
  \bibinfo{pages}{170801} (\bibinfo{year}{2015}).

\bibitem[{\citenamefont{Chaves et~al.}(2013)\citenamefont{Chaves, Brask,
  Markiewicz, {Ko{\l}ody\ifmmode \acute{n}\else {\'n}\fi{}ski}, and
  Ac{\'i}n}}]{PhysRevLett.111.120401}
\bibinfo{author}{\bibfnamefont{R.}~\bibnamefont{Chaves}},
  \bibinfo{author}{\bibfnamefont{J.~B.} \bibnamefont{Brask}},
  \bibinfo{author}{\bibfnamefont{M.}~\bibnamefont{Markiewicz}},
  \bibinfo{author}{\bibfnamefont{J.}~\bibnamefont{{Ko{\l}ody\ifmmode
  \acute{n}\else {\'n}\fi{}ski}}}, \bibnamefont{and}
  \bibinfo{author}{\bibfnamefont{A.}~\bibnamefont{Ac{\'i}n}},
  \bibinfo{journal}{Phys. Rev. Lett.} \textbf{\bibinfo{volume}{111}},
  \bibinfo{pages}{120401} (\bibinfo{year}{2013}).

\bibitem[{\citenamefont{Brask et~al.}(2015)\citenamefont{Brask, Chaves, and
  {Ko{\l}ody\ifmmode \acute{n}\else {\'n}\fi{}ski}}}]{PhysRevX.5.031010}
\bibinfo{author}{\bibfnamefont{J.~B.} \bibnamefont{Brask}},
  \bibinfo{author}{\bibfnamefont{R.}~\bibnamefont{Chaves}}, \bibnamefont{and}
  \bibinfo{author}{\bibfnamefont{J.}~\bibnamefont{{Ko{\l}ody\ifmmode
  \acute{n}\else {\'n}\fi{}ski}}}, \bibinfo{journal}{Phys. Rev. X}
  \textbf{\bibinfo{volume}{5}}, \bibinfo{pages}{031010} (\bibinfo{year}{2015}).

\bibitem[{\citenamefont{Kessler et~al.}(2014)\citenamefont{Kessler, Lovchinsky,
  Sushkov, and Lukin}}]{PhysRevLett.112.150802}
\bibinfo{author}{\bibfnamefont{E.~M.} \bibnamefont{Kessler}},
  \bibinfo{author}{\bibfnamefont{I.}~\bibnamefont{Lovchinsky}},
  \bibinfo{author}{\bibfnamefont{A.~O.} \bibnamefont{Sushkov}},
  \bibnamefont{and} \bibinfo{author}{\bibfnamefont{M.~D.} \bibnamefont{Lukin}},
  \bibinfo{journal}{Phys. Rev. Lett.} \textbf{\bibinfo{volume}{112}},
  \bibinfo{pages}{150802} (\bibinfo{year}{2014}).

\bibitem[{\citenamefont{D{\"u}r et~al.}(2014)\citenamefont{D{\"u}r,
  Skotiniotis, Fr{\"o}wis, and Kraus}}]{PhysRevLett.112.080801}
\bibinfo{author}{\bibfnamefont{W.}~\bibnamefont{D{\"u}r}},
  \bibinfo{author}{\bibfnamefont{M.}~\bibnamefont{Skotiniotis}},
  \bibinfo{author}{\bibfnamefont{F.}~\bibnamefont{Fr{\"o}wis}},
  \bibnamefont{and} \bibinfo{author}{\bibfnamefont{B.}~\bibnamefont{Kraus}},
  \bibinfo{journal}{Phys. Rev. Lett.} \textbf{\bibinfo{volume}{112}},
  \bibinfo{pages}{080801} (\bibinfo{year}{2014}).

\bibitem[{\citenamefont{Monz et~al.}(2011)\citenamefont{Monz, Schindler,
  Barreiro, Chwalla, Nigg, Coish, Harlander, H{\"a}nsel, Hennrich, and
  Blatt}}]{PhysRevLett.106.130506}
\bibinfo{author}{\bibfnamefont{T.}~\bibnamefont{Monz}},
  \bibinfo{author}{\bibfnamefont{P.}~\bibnamefont{Schindler}},
  \bibinfo{author}{\bibfnamefont{J.~T.} \bibnamefont{Barreiro}},
  \bibinfo{author}{\bibfnamefont{M.}~\bibnamefont{Chwalla}},
  \bibinfo{author}{\bibfnamefont{D.}~\bibnamefont{Nigg}},
  \bibinfo{author}{\bibfnamefont{W.~A.} \bibnamefont{Coish}},
  \bibinfo{author}{\bibfnamefont{M.}~\bibnamefont{Harlander}},
  \bibinfo{author}{\bibfnamefont{W.}~\bibnamefont{H{\"a}nsel}},
  \bibinfo{author}{\bibfnamefont{M.}~\bibnamefont{Hennrich}}, \bibnamefont{and}
  \bibinfo{author}{\bibfnamefont{R.}~\bibnamefont{Blatt}},
  \bibinfo{journal}{Phys. Rev. Lett.} \textbf{\bibinfo{volume}{106}},
  \bibinfo{pages}{130506} (\bibinfo{year}{2011}).

\bibitem[{\citenamefont{Micadei et~al.}(2015)\citenamefont{Micadei, Rowlands,
  Pollock, C{\'{e}}leri, Serra, and Modi}}]{micadei}
\bibinfo{author}{\bibfnamefont{K.}~\bibnamefont{Micadei}},
  \bibinfo{author}{\bibfnamefont{D.~A.} \bibnamefont{Rowlands}},
  \bibinfo{author}{\bibfnamefont{F.~A.} \bibnamefont{Pollock}},
  \bibinfo{author}{\bibfnamefont{L.~C.} \bibnamefont{C{\'{e}}leri}},
  \bibinfo{author}{\bibfnamefont{R.~M.} \bibnamefont{Serra}}, \bibnamefont{and}
  \bibinfo{author}{\bibfnamefont{K.}~\bibnamefont{Modi}}, \bibinfo{journal}{New
  J. Phys.} \textbf{\bibinfo{volume}{17}}, \bibinfo{pages}{023057}
  (\bibinfo{year}{2015}).

\bibitem[{\citenamefont{Maccone}(2013)}]{PhysRevA.88.042109}
\bibinfo{author}{\bibfnamefont{L.}~\bibnamefont{Maccone}},
  \bibinfo{journal}{Phys. Rev. A} \textbf{\bibinfo{volume}{88}},
  \bibinfo{pages}{042109} (\bibinfo{year}{2013}).

\bibitem[{\citenamefont{Baumgratz et~al.}(2014)\citenamefont{Baumgratz, Cramer,
  and Plenio}}]{Baumgratz2014}
\bibinfo{author}{\bibfnamefont{T.}~\bibnamefont{Baumgratz}},
  \bibinfo{author}{\bibfnamefont{M.}~\bibnamefont{Cramer}}, \bibnamefont{and}
  \bibinfo{author}{\bibfnamefont{M.~B.} \bibnamefont{Plenio}},
  \bibinfo{journal}{Phys. Rev. Lett.} \textbf{\bibinfo{volume}{113}},
  \bibinfo{pages}{140401} (\bibinfo{year}{2014}).

\bibitem[{\citenamefont{Marvian and Spekkens}(2014)}]{Marvian2014}
\bibinfo{author}{\bibfnamefont{I.}~\bibnamefont{Marvian}} \bibnamefont{and}
  \bibinfo{author}{\bibfnamefont{R.~W.} \bibnamefont{Spekkens}},
  \bibinfo{journal}{Nat. Commun.} \textbf{\bibinfo{volume}{5}},
  \bibinfo{pages}{3821} (\bibinfo{year}{2014}).

\bibitem[{\citenamefont{Marvian and Spekkens}(2016)}]{Marvian2016}
\bibinfo{author}{\bibfnamefont{I.}~\bibnamefont{Marvian}} \bibnamefont{and}
  \bibinfo{author}{\bibfnamefont{R.~W.} \bibnamefont{Spekkens}},
  \bibinfo{journal}{arXiv preprint arXiv:1602.08049}  (\bibinfo{year}{2016}).

\bibitem[{\citenamefont{Napoli et~al.}(2016)\citenamefont{Napoli, Bromley,
  Cianciaruso, Piani, Johnston, and Adesso}}]{Robus}
\bibinfo{author}{\bibfnamefont{C.}~\bibnamefont{Napoli}},
  \bibinfo{author}{\bibfnamefont{T.~R.} \bibnamefont{Bromley}},
  \bibinfo{author}{\bibfnamefont{M.}~\bibnamefont{Cianciaruso}},
  \bibinfo{author}{\bibfnamefont{M.}~\bibnamefont{Piani}},
  \bibinfo{author}{\bibfnamefont{N.}~\bibnamefont{Johnston}}, \bibnamefont{and}
  \bibinfo{author}{\bibfnamefont{G.}~\bibnamefont{Adesso}},
  \bibinfo{journal}{Phys. Rev. Lett.} \textbf{\bibinfo{volume}{116}},
  \bibinfo{pages}{150502} (\bibinfo{year}{2016}).

\bibitem[{\citenamefont{Higgins et~al.}(2007)\citenamefont{Higgins, Berry,
  Bartlett, Wiseman, and Pryde}}]{higgins2007entanglement}
\bibinfo{author}{\bibfnamefont{B.~L.} \bibnamefont{Higgins}},
  \bibinfo{author}{\bibfnamefont{D.~W.} \bibnamefont{Berry}},
  \bibinfo{author}{\bibfnamefont{S.~D.} \bibnamefont{Bartlett}},
  \bibinfo{author}{\bibfnamefont{H.~M.} \bibnamefont{Wiseman}},
  \bibnamefont{and} \bibinfo{author}{\bibfnamefont{G.~J.} \bibnamefont{Pryde}},
  \bibinfo{journal}{Nature} \textbf{\bibinfo{volume}{450}},
  \bibinfo{pages}{393} (\bibinfo{year}{2007}).

\bibitem[{\citenamefont{Holevo}(2011)}]{holevo2011probabilistic}
\bibinfo{author}{\bibfnamefont{A.~S.} \bibnamefont{Holevo}},
  \emph{\bibinfo{title}{{Probabilistic and Statistical Aspects of Quantum
  Theory}}} (\bibinfo{publisher}{Edizioni della Normale, Pisa},
  \bibinfo{year}{2011}), \bibinfo{edition}{1st} ed.

\bibitem[{\citenamefont{Perez-Garcia et~al.}(2006)\citenamefont{Perez-Garcia,
  Wolf, Petz, and Ruskai}}]{perez2006contractivity}
\bibinfo{author}{\bibfnamefont{D.}~\bibnamefont{Perez-Garcia}},
  \bibinfo{author}{\bibfnamefont{M.~M.} \bibnamefont{Wolf}},
  \bibinfo{author}{\bibfnamefont{D.}~\bibnamefont{Petz}}, \bibnamefont{and}
  \bibinfo{author}{\bibfnamefont{M.~B.} \bibnamefont{Ruskai}},
  \bibinfo{journal}{Journal of Mathematical Physics}
  \textbf{\bibinfo{volume}{47}}, \bibinfo{pages}{083506}
  (\bibinfo{year}{2006}).

\bibitem[{\citenamefont{Nielsen and Chuang}(2000)}]{nielsen2000}
\bibinfo{author}{\bibfnamefont{M.~A.} \bibnamefont{Nielsen}} \bibnamefont{and}
  \bibinfo{author}{\bibfnamefont{I.~L.} \bibnamefont{Chuang}},
  \emph{\bibinfo{title}{{Quantum computation and quantum information}}}
  (\bibinfo{publisher}{Cambridge University Press}, \bibinfo{year}{2000}).

\bibitem[{\citenamefont{Bengtsson and
  Zyczkowski}(2006)}]{bengtsson2006geometry}
\bibinfo{author}{\bibfnamefont{I.}~\bibnamefont{Bengtsson}} \bibnamefont{and}
  \bibinfo{author}{\bibfnamefont{K.}~\bibnamefont{Zyczkowski}},
  \emph{\bibinfo{title}{{Geometry of quantum states: an introduction to quantum
  entanglement}}} (\bibinfo{publisher}{Cambridge University Press},
  \bibinfo{year}{2006}).

\bibitem[{\citenamefont{Braun et~al.}(2014)\citenamefont{Braun, Giraud, Nechita
  et~al.}}]{braun2014universal}
\bibinfo{author}{\bibfnamefont{D.}~\bibnamefont{Braun}},
  \bibinfo{author}{\bibfnamefont{O.}~\bibnamefont{Giraud}},
  \bibinfo{author}{\bibfnamefont{I.}~\bibnamefont{Nechita}},
  \bibnamefont{et~al.}, \bibinfo{journal}{J. Phys. A: Math. Theor.}
  \textbf{\bibinfo{volume}{47}}, \bibinfo{pages}{135302}
  (\bibinfo{year}{2014}).

\bibitem[{\citenamefont{Braunstein and Caves}(1994)}]{PhysRevLett.72.3439}
\bibinfo{author}{\bibfnamefont{S.~L.} \bibnamefont{Braunstein}}
  \bibnamefont{and} \bibinfo{author}{\bibfnamefont{C.~M.} \bibnamefont{Caves}},
  \bibinfo{journal}{Phys. Rev. Lett.} \textbf{\bibinfo{volume}{72}},
  \bibinfo{pages}{3439} (\bibinfo{year}{1994}).

\bibitem[{\citenamefont{Barndorff-Nielsen and
  Gill}(2000)}]{barndorff2000fisher}
\bibinfo{author}{\bibfnamefont{O.}~\bibnamefont{Barndorff-Nielsen}}
  \bibnamefont{and} \bibinfo{author}{\bibfnamefont{R.}~\bibnamefont{Gill}},
  \bibinfo{journal}{J. Phys. A: Math. Gen.} \textbf{\bibinfo{volume}{33}},
  \bibinfo{pages}{4481} (\bibinfo{year}{2000}).

\bibitem[{\citenamefont{Gill and Levit}(1995)}]{gilltrees}
\bibinfo{author}{\bibfnamefont{R.~D.} \bibnamefont{Gill}} \bibnamefont{and}
  \bibinfo{author}{\bibfnamefont{B.~Y.} \bibnamefont{Levit}},
  \bibinfo{journal}{Bernoulli} \textbf{\bibinfo{volume}{1}},
  \bibinfo{pages}{59} (\bibinfo{year}{1995}).

\bibitem[{\citenamefont{Tsang}(2012)}]{mankey}
\bibinfo{author}{\bibfnamefont{M.}~\bibnamefont{Tsang}},
  \bibinfo{journal}{Phys. Rev. Lett.} \textbf{\bibinfo{volume}{108}},
  \bibinfo{pages}{230401} (\bibinfo{year}{2012}).

\bibitem[{\citenamefont{Sekatski et~al.}(2016)\citenamefont{Sekatski,
  Skotiniotis, {Ko\l{}ody\'{n}ski}, and D\"ur}}]{kolo2016}
\bibinfo{author}{\bibfnamefont{P.}~\bibnamefont{Sekatski}},
  \bibinfo{author}{\bibfnamefont{M.}~\bibnamefont{Skotiniotis}},
  \bibinfo{author}{\bibfnamefont{J.}~\bibnamefont{{Ko\l{}ody\'{n}ski}}},
  \bibnamefont{and} \bibinfo{author}{\bibfnamefont{W.}~\bibnamefont{D\"ur}},
  \bibinfo{journal}{arXiv preprint arXiv:1603.08944}  (\bibinfo{year}{2016}).

\bibitem[{\citenamefont{Paris}(2009)}]{paris2009quantum}
\bibinfo{author}{\bibfnamefont{M.~G.~A.} \bibnamefont{Paris}},
  \bibinfo{journal}{Int. J. Quant. Inf.} \textbf{\bibinfo{volume}{7}},
  \bibinfo{pages}{125} (\bibinfo{year}{2009}).

\bibitem[{\citenamefont{Jing et~al.}(2014)\citenamefont{Jing, Xiao-Xing, Wei,
  and Xiao-Guang}}]{0253-6102-61-1-08}
\bibinfo{author}{\bibfnamefont{L.}~\bibnamefont{Jing}},
  \bibinfo{author}{\bibfnamefont{J.}~\bibnamefont{Xiao-Xing}},
  \bibinfo{author}{\bibfnamefont{Z.}~\bibnamefont{Wei}}, \bibnamefont{and}
  \bibinfo{author}{\bibfnamefont{W.}~\bibnamefont{Xiao-Guang}},
  \bibinfo{journal}{Comm. Theor. Phys.} \textbf{\bibinfo{volume}{61}},
  \bibinfo{pages}{115} (\bibinfo{year}{2014}).

\bibitem[{\citenamefont{Demkowicz-Dobrza{\'n}ski and
  Maccone}(2014)}]{demkowicz2014using}
\bibinfo{author}{\bibfnamefont{R.}~\bibnamefont{Demkowicz-Dobrza{\'n}ski}}
  \bibnamefont{and} \bibinfo{author}{\bibfnamefont{L.}~\bibnamefont{Maccone}},
  \bibinfo{journal}{Phys. Rev. Lett.} \textbf{\bibinfo{volume}{113}},
  \bibinfo{pages}{250801} (\bibinfo{year}{2014}).

\bibitem[{\citenamefont{Yousefjani et~al.}(2016)\citenamefont{Yousefjani,
  Salimi, and Khorashad}}]{yousefjani2016framework}
\bibinfo{author}{\bibfnamefont{R.}~\bibnamefont{Yousefjani}},
  \bibinfo{author}{\bibfnamefont{S.}~\bibnamefont{Salimi}}, \bibnamefont{and}
  \bibinfo{author}{\bibfnamefont{A.}~\bibnamefont{Khorashad}},
  \bibinfo{journal}{arXiv preprint arXiv:1602.01691}  (\bibinfo{year}{2016}).

\bibitem[{\citenamefont{Alipour et~al.}(2014)\citenamefont{Alipour, Mehboudi,
  and Rezakhani}}]{alipour2014quantum}
\bibinfo{author}{\bibfnamefont{S.}~\bibnamefont{Alipour}},
  \bibinfo{author}{\bibfnamefont{M.}~\bibnamefont{Mehboudi}}, \bibnamefont{and}
  \bibinfo{author}{\bibfnamefont{A.}~\bibnamefont{Rezakhani}},
  \bibinfo{journal}{Phys. Rev. Lett.} \textbf{\bibinfo{volume}{112}},
  \bibinfo{pages}{120405} (\bibinfo{year}{2014}).

\bibitem[{\citenamefont{Karpi{\'n}ski et~al.}(2008)\citenamefont{Karpi{\'n}ski,
  Radzewicz, and Banaszek}}]{karpinski2008fiber}
\bibinfo{author}{\bibfnamefont{M.}~\bibnamefont{Karpi{\'n}ski}},
  \bibinfo{author}{\bibfnamefont{C.}~\bibnamefont{Radzewicz}},
  \bibnamefont{and} \bibinfo{author}{\bibfnamefont{K.}~\bibnamefont{Banaszek}},
  \bibinfo{journal}{JOSA B} \textbf{\bibinfo{volume}{25}}, \bibinfo{pages}{668}
  (\bibinfo{year}{2008}).

\bibitem[{\citenamefont{Fisher}(1953)}]{fisher1953dispersion}
\bibinfo{author}{\bibfnamefont{R.}~\bibnamefont{Fisher}},
  \bibinfo{journal}{Proc. R. Soc. A.} \textbf{\bibinfo{volume}{217}}, \bibinfo{pages}{295}
  (\bibinfo{year}{1953}).

\bibitem[{\citenamefont{Girolami et~al.}(2014)\citenamefont{Girolami, Souza,
  Giovannetti, Tufarelli, Filgueiras, Sarthour, Soares-Pinto, Oliveira, and
  Adesso}}]{interpower}
\bibinfo{author}{\bibfnamefont{D.}~\bibnamefont{Girolami}},
  \bibinfo{author}{\bibfnamefont{A.~M.} \bibnamefont{Souza}},
  \bibinfo{author}{\bibfnamefont{V.}~\bibnamefont{Giovannetti}},
  \bibinfo{author}{\bibfnamefont{T.}~\bibnamefont{Tufarelli}},
  \bibinfo{author}{\bibfnamefont{J.~G.} \bibnamefont{Filgueiras}},
  \bibinfo{author}{\bibfnamefont{R.~S.} \bibnamefont{Sarthour}},
  \bibinfo{author}{\bibfnamefont{D.~O.} \bibnamefont{Soares-Pinto}},
  \bibinfo{author}{\bibfnamefont{I.~S.} \bibnamefont{Oliveira}},
  \bibnamefont{and} \bibinfo{author}{\bibfnamefont{G.}~\bibnamefont{Adesso}},
  \bibinfo{journal}{Phys. Rev. Lett.} \textbf{\bibinfo{volume}{112}},
  \bibinfo{pages}{210401} (\bibinfo{year}{2014}).

\bibitem[{\citenamefont{Farace et~al.}(2016)\citenamefont{Farace, {De
  Pasquale}, Adesso, and Giovannetti}}]{avsk}
\bibinfo{author}{\bibfnamefont{A.}~\bibnamefont{Farace}},
  \bibinfo{author}{\bibfnamefont{A.}~\bibnamefont{{De Pasquale}}},
  \bibinfo{author}{\bibfnamefont{G.}~\bibnamefont{Adesso}}, \bibnamefont{and}
  \bibinfo{author}{\bibfnamefont{V.}~\bibnamefont{Giovannetti}},
  \bibinfo{journal}{New J. Phys.} \textbf{\bibinfo{volume}{18}},
  \bibinfo{pages}{013049} (\bibinfo{year}{2016}).

\bibitem[{\citenamefont{Greenberger et~al.}(1989)\citenamefont{Greenberger,
  Horne, and Zeilinger}}]{greenberger1989going}
\bibinfo{author}{\bibfnamefont{D.~M.} \bibnamefont{Greenberger}},
  \bibinfo{author}{\bibfnamefont{M.~A.} \bibnamefont{Horne}}, \bibnamefont{and}
  \bibinfo{author}{\bibfnamefont{A.}~\bibnamefont{Zeilinger}}, in
  \emph{\bibinfo{booktitle}{{Bell{\rq}s theorem, quantum theory and conceptions
  of the universe}}} (\bibinfo{publisher}{Springer}, \bibinfo{year}{1989}), pp.
  \bibinfo{pages}{69--72}.

\bibitem[{\citenamefont{Jarzyna and Demkowicz-Dobrza{\'n}ski}(2013)}]{metromps}
\bibinfo{author}{\bibfnamefont{M.}~\bibnamefont{Jarzyna}} \bibnamefont{and}
  \bibinfo{author}{\bibfnamefont{R.}~\bibnamefont{Demkowicz-Dobrza{\'n}ski}},
  \bibinfo{journal}{Phys. Rev. Lett.} \textbf{\bibinfo{volume}{110}},
  \bibinfo{pages}{240405} (\bibinfo{year}{2013}).

\bibitem[{\citenamefont{Baumgratz and Datta}(2016)}]{Datta2016}
\bibinfo{author}{\bibfnamefont{T.}~\bibnamefont{Baumgratz}} \bibnamefont{and}
  \bibinfo{author}{\bibfnamefont{A.}~\bibnamefont{Datta}},
  \bibinfo{journal}{Phys. Rev. Lett.} \textbf{\bibinfo{volume}{116}},
  \bibinfo{pages}{030801} (\bibinfo{year}{2016}).

\bibitem[{\citenamefont{Zwierz et~al.}(2010)\citenamefont{Zwierz,
  P\'erez-Delgado, and Kok}}]{Kok}
\bibinfo{author}{\bibfnamefont{M.}~\bibnamefont{Zwierz}},
  \bibinfo{author}{\bibfnamefont{C.~A.} \bibnamefont{P\'erez-Delgado}},
  \bibnamefont{and} \bibinfo{author}{\bibfnamefont{P.}~\bibnamefont{Kok}},
  \bibinfo{journal}{Phys. Rev. Lett.} \textbf{\bibinfo{volume}{105}},
  \bibinfo{pages}{180402} (\bibinfo{year}{2010}).

\bibitem[{\citenamefont{del Rio et~al.}(2015)\citenamefont{del Rio, Kraemer,
  and Renner}}]{del2015resource}
\bibinfo{author}{\bibfnamefont{L.}~\bibnamefont{del Rio}},
  \bibinfo{author}{\bibfnamefont{L.}~\bibnamefont{Kraemer}}, \bibnamefont{and}
  \bibinfo{author}{\bibfnamefont{R.}~\bibnamefont{Renner}},
  \bibinfo{journal}{arXiv preprint arXiv:1511.08818}  (\bibinfo{year}{2015}).

\bibitem[{\citenamefont{Breuer and Petruccione}(2002)}]{breuer2002theory}
\bibinfo{author}{\bibfnamefont{H.}~\bibnamefont{Breuer}} \bibnamefont{and}
  \bibinfo{author}{\bibfnamefont{F.}~\bibnamefont{Petruccione}},
  \emph{\bibinfo{title}{{The Theory of Open Quantum Systems}}}
  (\bibinfo{publisher}{Oxford University Press, USA}, \bibinfo{year}{2002}).

\end{thebibliography}

\end{document}